\documentclass[preprintnumbers]{revtex4}
\usepackage{amsmath}
\usepackage{graphicx}

\begin{document}

\title{Stability and Causality in relativistic dissipative hydrodynamics}
\author{G. S. Denicol, T. Kodama, T. Koide and Ph. Mota}
\affiliation{Instituto de F\'{\i}sica, Universidade Federal do Rio de Janeiro, C. P.
68528, 21945-970, Rio de Janeiro, Brazil}

\begin{abstract}
The stability and causality of the Landau-Lifshitz theory and the
Israel-Stewart type causal dissipative hydrodynamics are discussed. We show
that the problem of acausality and instability are correlated in
relativistic dissipative hydrodynamics and instability is induced by
acausality. We further discuss the stability of the scaling solution. The
scaling solution of the causal dissipative hydrodynamics can be unstable
against inhomogeneous perturbations.
\end{abstract}

\pacs{47.10.-g,25.75.-q}
\maketitle

\section{Introduction}

Presently the hydrodynamical approach is known as one of
basic tools for the description of the collective aspects of the
relativistic heavy-ion collisions.
However, instead of extensive
analyses of experimental data based on this approach \cite{review}, 
the study of the effect of dissipation is yet poorly explored 
\cite{rhydro,dkkm1,dkkm2}.

Till now, two basically different approaches to describe
relativistic dissipative hydrodynamics are commonly employed; 
one is the relativistic extension of Navier-Stokes theory introduced by
Landau-Lifshitz (LL) \cite{LL} and Eckart \cite{eckart}, and the other is
the causal dissipative (CD) hydrodynamics or second order theory, where the
version due to Israel-Stewart \cite{IS} is most popularly known. 
The relativistic Navier-Stokes theory (NS) is usually considered in the
literature as a natural covariant generalization of the Navier-Stokes
equation and in contrast to CD hydrodynamics, it is also refereed to
as the first order theory. Authors who applies the first order
theory argue that the second order theory is not yet completely established
and the difference of two approaches would be not significant, since the
effect of viscosity is the measure of the deviation from the local
thermodynamical equilibrium, and should be small where the hydrodynamical
approach is meaningful. However, a crucial point of the reason why
the second order approach should be explored is that the propagation speed
in the first order theory is infinite and it does not satisfy the
relativistic causality. It is also known that the first order theory leads
to dynamical instabilities. These aspects are not only the question of
conceptions but also related to serious practical problems and deserve a
more detailed analysis. 

One way to solve the problem of acausality in a first order theory
is introduce, for example, a memory effect with a finite relaxation time 
\cite{dkkm1}. 
In this way, we can derive a CD hydrodynamics from the LL theory. 
There are several different approaches to derive the CD
hydrodynamics \cite{IS,dkkm1,jou,muller,carter,OG}. 
In this work, we consider the Israel-Stewart type CD hydrodynamics.

A crucial point is whether a relativistic dissipative hydrodynamics
is stable or not. If a theory is unstable, it will be very difficult to
extract any physically meaningful results from it. The analysis of the
stability of relativistic dissipative hydrodynamics has been extensively
studied by Hiscock and his collaborators \cite{his1,his2,his3,his4,his5,his6}%
. Their conclusions are summarized as follows. 1) The Eckart theory is
unstable for the linear perturbation around the hydrostatic states 
whereas the LL theory and the IS theory (of the Landau frame) are stable 
\cite{his2,his3}. 2) The LL theory is shown to be unstable for the linear
perturbation around hydrostatic states in a general frame where the
fluid is Lorentz boosted \cite{his2}.

Another important analysis was implemented by Kouno et al \cite{kouno}. They
discussed the linear perturbation around the scaling solution of the LL
theory and found that the scaling solution of the LL theory can be unstable.
A similar discussion is repeated by \cite{gio} using another equation of
state.

The purpose of this paper is to complement these discussions. We will, in
particular, focus on two subjects. One is the relation between causality and
stability. In general, the concept of causality and stability are
independent, but in relativistic systems, as we will see, they 
are closely related. To show this, we discuss the
stability from Lorentz boosted frames and show that instability is induced
because of acausality.

The other is the stability around the scaling solution in the CD
hydrodynamics. In the LL theory, the scaling solution can be unstable,
although it is always stable when the Reynolds number is larger than one.
The stability around the scaling solution has not yet been discussed in the
CD hydrodynamics. The applicability of the scaling ansatz is not obvious in
the CD hydrodynamics, because, as we will see later, the numerical
calculation of the CD hydrodynamics shows a kind of non-periodic
oscillations in the central rapidity region.

In this paper, we restrict ourselves to the discussion on 1+1
dimensional motion of massless ideal gas (to be specific, an ideal three
flavor massless QGP gas), for simplicity. This paper is organized as
follows. In section \ref{chap:2}, we discuss the stability and causality
around the hydrostatic states. The result of this section has already been
shown in \cite{his2,his3,dkkm2}. To establish the relation between causality
and stability, we discuss the stability of the hydrostatic state from
Lorentz boosted frames in section \ref{chap:3}. In section \ref{chap:4}, the
stability around the scaling solution is discussed. Section \ref{chap:5} is
devoted to concluding remarks.

\section{Linear perturbation around the hydrostatic state}

\label{chap:2}

Before discussing the relation between causality and stability, we will
discuss first the stability near the hydrostatic state following \cite%
{his2,his3,dkkm2}.

In the CD hydrodynamics of 1+1 dimensional systems, the equations are given
by 
\begin{eqnarray}
\partial _{\nu }T^{\mu \nu } &=&0, \\
\tau _{R}\frac{d}{d\tau }\Pi +\Pi &=&-\zeta \partial _{\mu }u^{\mu }.
\end{eqnarray}%
Here, $\Pi $ is the bulk viscosity and $T^{\mu \nu }$ is the energy-momentum
tensor defined by 
\begin{eqnarray}
T^{\mu \nu }=(\varepsilon +P+\Pi )u^{\mu }u^{\nu }-(P+\Pi )g^{\mu \nu },
\end{eqnarray}%
where $\varepsilon $, $P$ and $u^{\mu }$ are the energy density, pressure
and four-velocity of the fluid, respectively. 
For the massless QGP, we have $\varepsilon =3P,$ and $\varepsilon +P=Ts,$%
where $T$ and $s$ are the temperature and
entropy density, respectively. The fluid velocity is determined
from the energy-momentum tensor following the definition of Landau-Lifshitz 
\cite{LL}. It should be noted that the CD hydrodynamics is reduced to the LL
theory in the limit of the vanishing relaxation time $\tau _{R}$.

For later convenience, we parametrize the velocity as follows, 
\begin{eqnarray}
u^{\mu }=(\cosh \theta ,\sinh \theta ).
\end{eqnarray}%
We adopt the following parametrization of the bulk viscosity coefficient and
the relaxation time \cite{dkkm2}, 
\begin{eqnarray}
\zeta &=&as, \\
\tau _{R} &=&\frac{\zeta }{\varepsilon +P}b,  \label{eqn:def-tau}
\end{eqnarray}%
where $s$ is the entropy density. The parameters $a$ and $b$ characterize
respectively the magnitudes of the viscosity and the relaxation time \cite%
{footnote1}. As we will see later, the parameter $b$ should be smaller than $%
3/2$ to be consistent with causality \cite{dkkm1}.

We consider the plane wave perturbation around $\varepsilon_0$, $\theta_0$
and $\Pi_0$, 
\begin{eqnarray}
\varepsilon (t,x) &\approx & \varepsilon_0(t,x) + \epsilon
\varepsilon_1(t,x), \\
\theta(t,x) &\approx & \theta_0 (t,x)+ \epsilon \theta_1(t,x), \\
\Pi(t,x) &\approx & \Pi_0 (t,x)+ \epsilon \Pi_1(t,x),
\end{eqnarray}
with 
\begin{eqnarray}
\left( 
\begin{array}{c}
\varepsilon_1 \left( t,x\right) \\ 
\theta_1 \left( t,x\right) \\ 
\Pi_1 \left( t,x\right)%
\end{array}
\right) =e^{i\omega t-ikx}\left( 
\begin{array}{c}
\varepsilon_1 \\ 
\theta_1 \\ 
\Pi_1%
\end{array}
\right).
\end{eqnarray}
Here $\epsilon$ is a small expansion parameter. The two transport
coefficients are also expanded as 
\begin{eqnarray}
\zeta &\approx& \zeta_0 + \epsilon \delta \zeta, \\
\tau_R =&\approx& \tau_{R0} + \epsilon \delta \tau_R.
\end{eqnarray}

For the linear perturbation around the hydrostatic state, 
\begin{eqnarray}
\varepsilon _{0eqnarray*}=\mathrm{const},~~~~\theta _{0}=\Pi _{0}=0,
\end{eqnarray}%
the evolution equation of the linear perturbation is given by 
\begin{eqnarray}
A\left( 
\begin{array}{c}
\varepsilon _{1} \\ 
\theta _{1} \\ 
\Pi _{1}%
\end{array}%
\right) =0,
\end{eqnarray}%
where 
\begin{equation}
A=\left( 
\begin{array}{ccc}
i\omega & -ik(\varepsilon _{0}+P_{0}) & 0 \\ 
\alpha (-ik) & i\omega (\varepsilon _{0}+P_{0}) & -ik \\ 
0 & -ik\zeta _{0} & 1+\tau _{R0}i\omega%
\end{array}%
\right) ,
\end{equation}%
where $\alpha =\partial P/\partial \varepsilon =1/3$. To have non-trivial
solutions, the determinant of the matrix $A$ should vanish so that the
frequency $\omega $ has to satisfy the following dispersion relation, 
\begin{equation}
\omega ^{3}-\frac{i}{\tau _{R0}}\omega ^{2}-\left( \frac{\zeta _{0}}{\tau
_{R}}\frac{1}{\varepsilon _{0}+P_{0}}+\alpha \right) k^{2}\omega +i\frac{%
\alpha }{\tau _{R0}}k^{2}=0.  \label{eqn:disp}
\end{equation}%
The behaviors of the frequency characterizes the stability and the
propagation speed of the fluid.

\subsection{Landau-Lifshitz theory}

We consider the case of the LL theory by taking $\tau_R = 0$. Then the
solution of the dispersion relation (\ref{eqn:disp}) is analytically given
by 
\begin{eqnarray}
\omega = \frac{i\zeta_0}{2(\varepsilon_0 + P_0)} \pm \sqrt{ \alpha k^2 - 
\frac{\zeta_0^2}{4(\varepsilon + P_0)^2}k^4 }.
\end{eqnarray}
The real and imaginary parts of the frequency $\omega$ are shown in Fig. \ref%
{LL_0} for $a = 0.1$ and $T=200$ MeV, respectively. One can easily see that
the behavior of $\omega$ changes at the critical momentum $k_{c} = 2\sqrt{%
\alpha} (\varepsilon + P)/\zeta_0$; below the critical momentum, there are
two propagating modes, while they are changed to two non-propagating modes
above it.

We assume that the propagation speed of fluid is characterized by the group
velocity for propagating modes. Then, the causality of the theory is
determined by the behavior of the real parts of the frequencies. For the
small $k$, the propagation speed is given by 
\begin{eqnarray}
v = \frac{\partial \mathrm{Re}~\omega}{\partial k} \approx \sqrt{\alpha}.
\end{eqnarray}
This is nothing but the usual sound velocity, and the LL theory seems to be
consistent with causality. For the large $k$, however, the propagating modes
are changed to the non-propagating modes which show $k^2$ dependence. This
momentum dependence is the same behavior as that of the non-propagating mode
in diffusion processes, where the propagation speed is infinite. It is
considered that the behavior of the non-propagating mode is the origin of
acausality in the LL theory.

On the other hand, the stability of the theory is characterized by the
behaviors of the imaginary parts of the frequencies. One can easily see that
the two modes always have positive imaginary parts, and hence the LL theory
is stable under the linear perturbation around the hydrostatic states.

Note that this is different behavior from the Eckart theory, where the
theory is acausal and unstable even for the linear perturbation around the
hydrostatic state \cite{his2}.

\begin{figure}[tbp]
\begin{minipage}{.45\linewidth}
\includegraphics[scale=0.3]{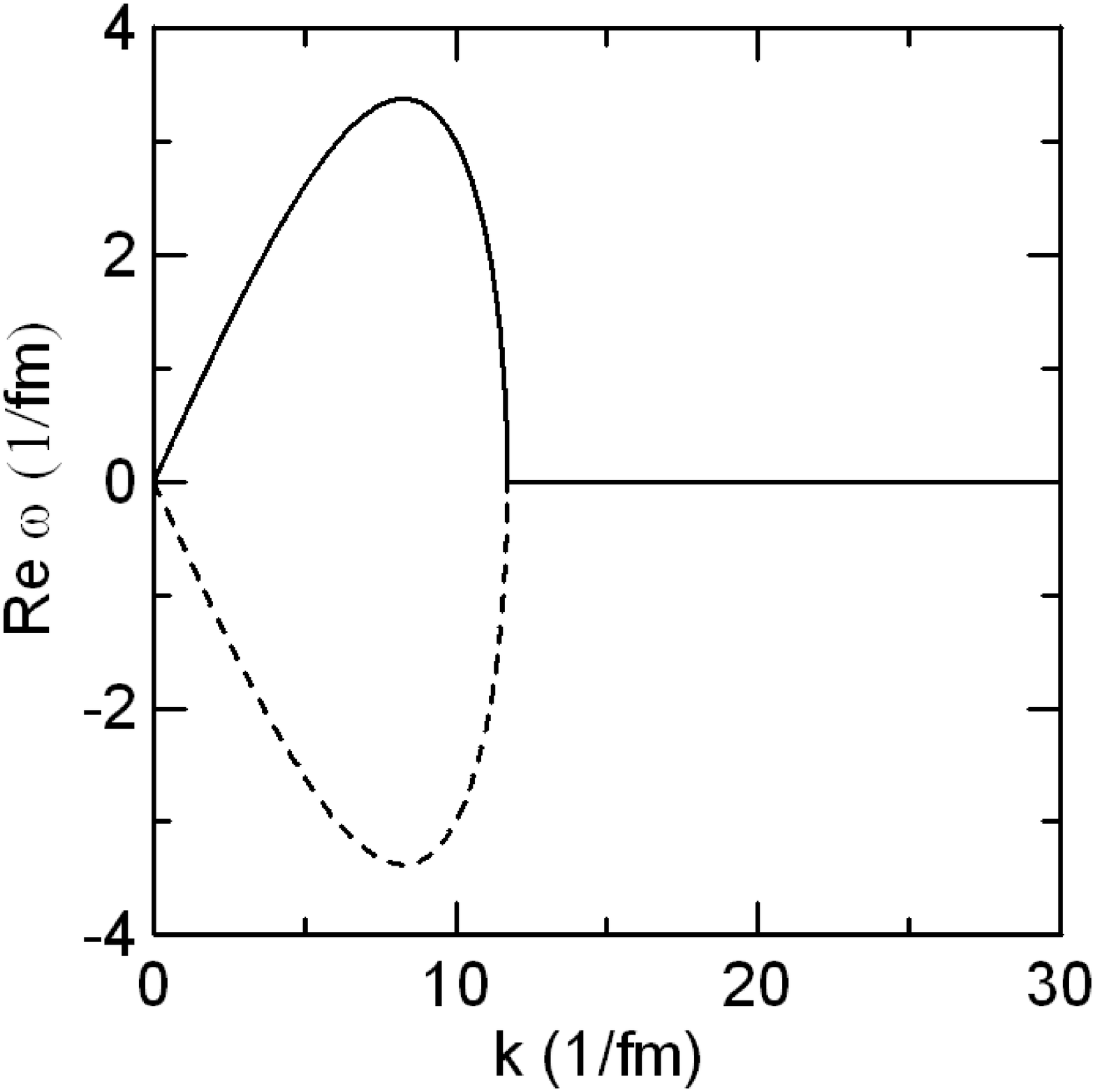}
\end{minipage}
\begin{minipage}{.45\linewidth}
\includegraphics[scale=0.3]{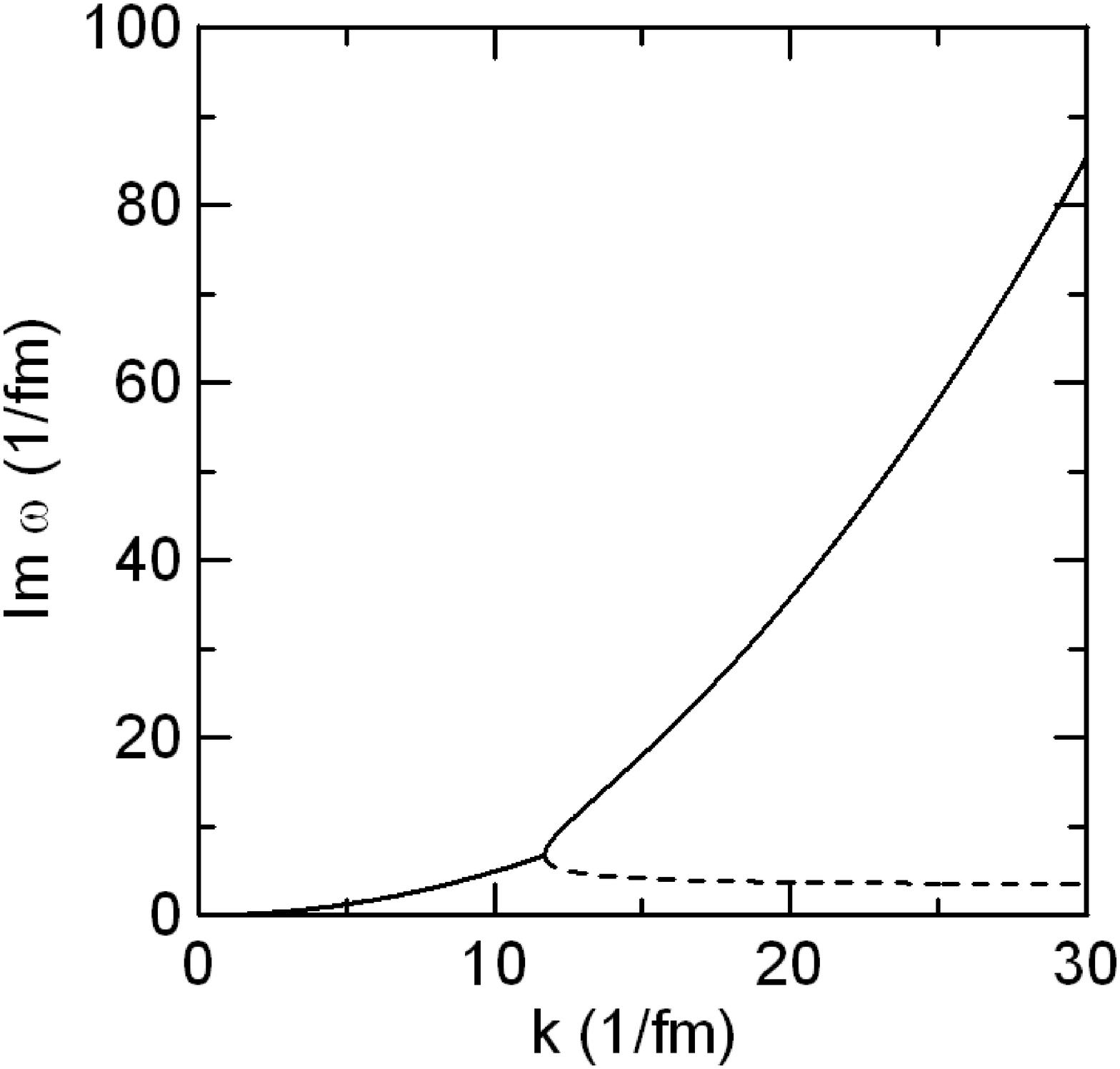}
\end{minipage}
\caption{The real (left panel) and imaginary (right panel) parts of the
frequency in the Landau-Lifshitz theory at the rest frame. The two
propagating modes (the solid and dotted lines) are changed to the
non-propagating modes at the critical $k_c$. Both of them have positive
values.}
\label{LL_0}
\end{figure}

\subsection{Causal dissipative hydrodynamics}

Different from the case of the LL theory, we can still obtain the
dispersion relation from Eq.(\ref{eqn:disp}), but the analytic form of the
solution becomes extremely complicated. However, in the large $k$ limit, we
have \cite{dkkm2} 
\begin{eqnarray}
\omega =\left\{ 
\begin{array}{c}
\pm k\sqrt{\frac{1}{b}+\alpha }+\frac{i}{2\tau _{R}(1+\alpha b)} \\ 
i\frac{\alpha b}{\tau _{R}(1+\alpha b)}%
\end{array}%
\right. , \label{eqn:Lk}
\end{eqnarray}%
while for the small $k$, 
\begin{eqnarray}
\omega =\left\{ 
\begin{array}{c}
\pm k\frac{\sqrt{\alpha -(\alpha +1/(4b))\tau _{R}^{2}k^{2}/b}}{1-\tau
_{R}^{2}k^{2}/b}+\frac{ik^{2}\tau _{R}}{2(b-\tau _{R}^{2}k^{2})} \\ 
i/\tau _{R}%
\end{array}%
\right. . \label{eqn:Sk}
\end{eqnarray}

In this case, there are three modes; two of them are propagating modes and
the remaining one is a non-propagating mode. From Eqs.(\ref{eqn:Lk}) and (%
\ref{eqn:Sk}), we can see that, for the small k, the group velocity of the
propagating modes reduces to that of the ideal one, $\sqrt{\alpha}$, like
the LL theory. On the other hand, for the large k, the group velocity is
given by 
\begin{eqnarray}
v_{c}=\sqrt{1/b + \alpha}.  \label{eqn:prop_cd}
\end{eqnarray}
That is, the group velocity is affected by the bulk viscosity. This gives
the maximum propagation speed of this fluid.

\begin{figure}[tbp]
\begin{minipage}{.45\linewidth}
\includegraphics[scale=0.3]{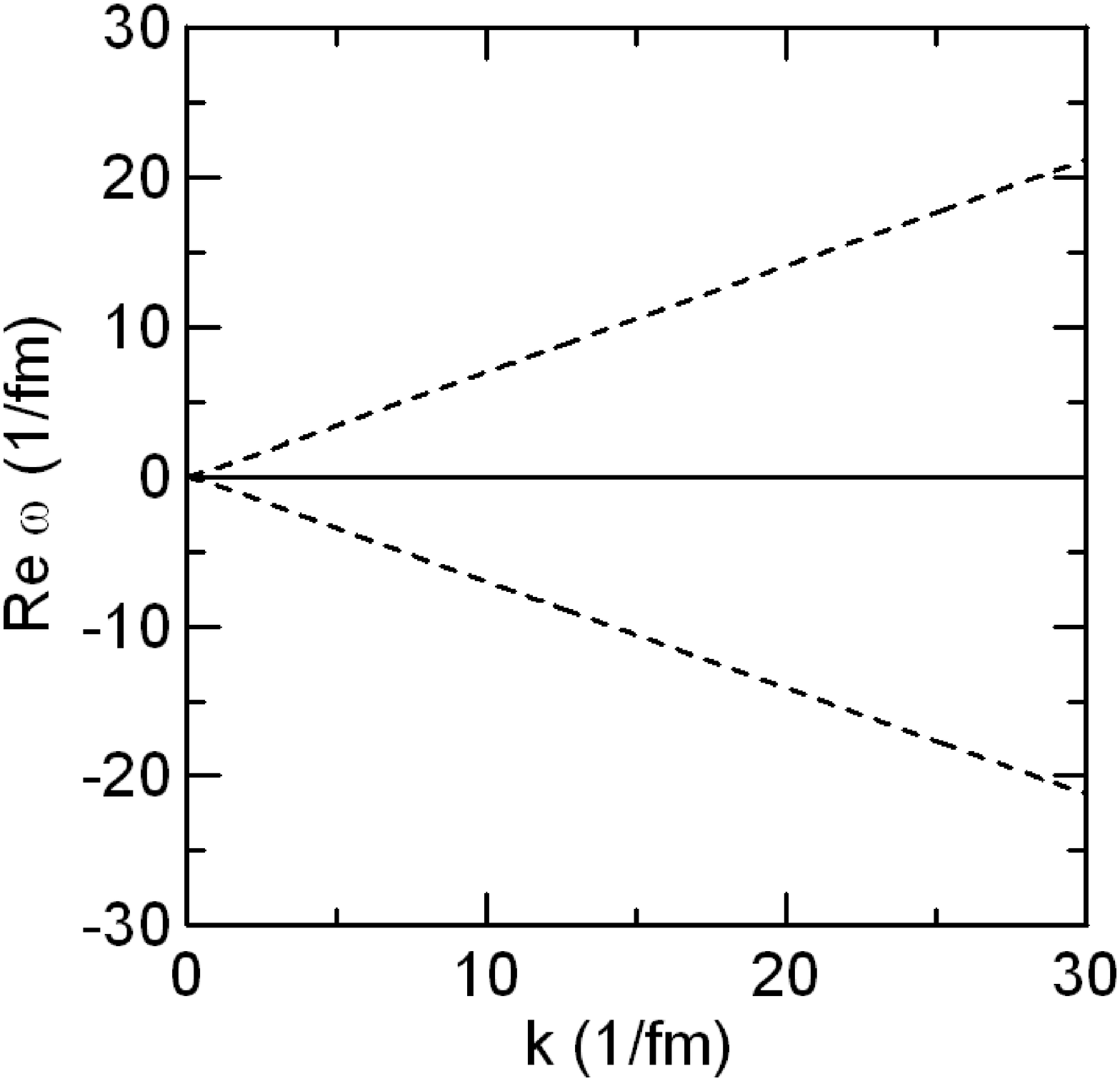}
\end{minipage}
\begin{minipage}{.45\linewidth}
\includegraphics[scale=0.3]{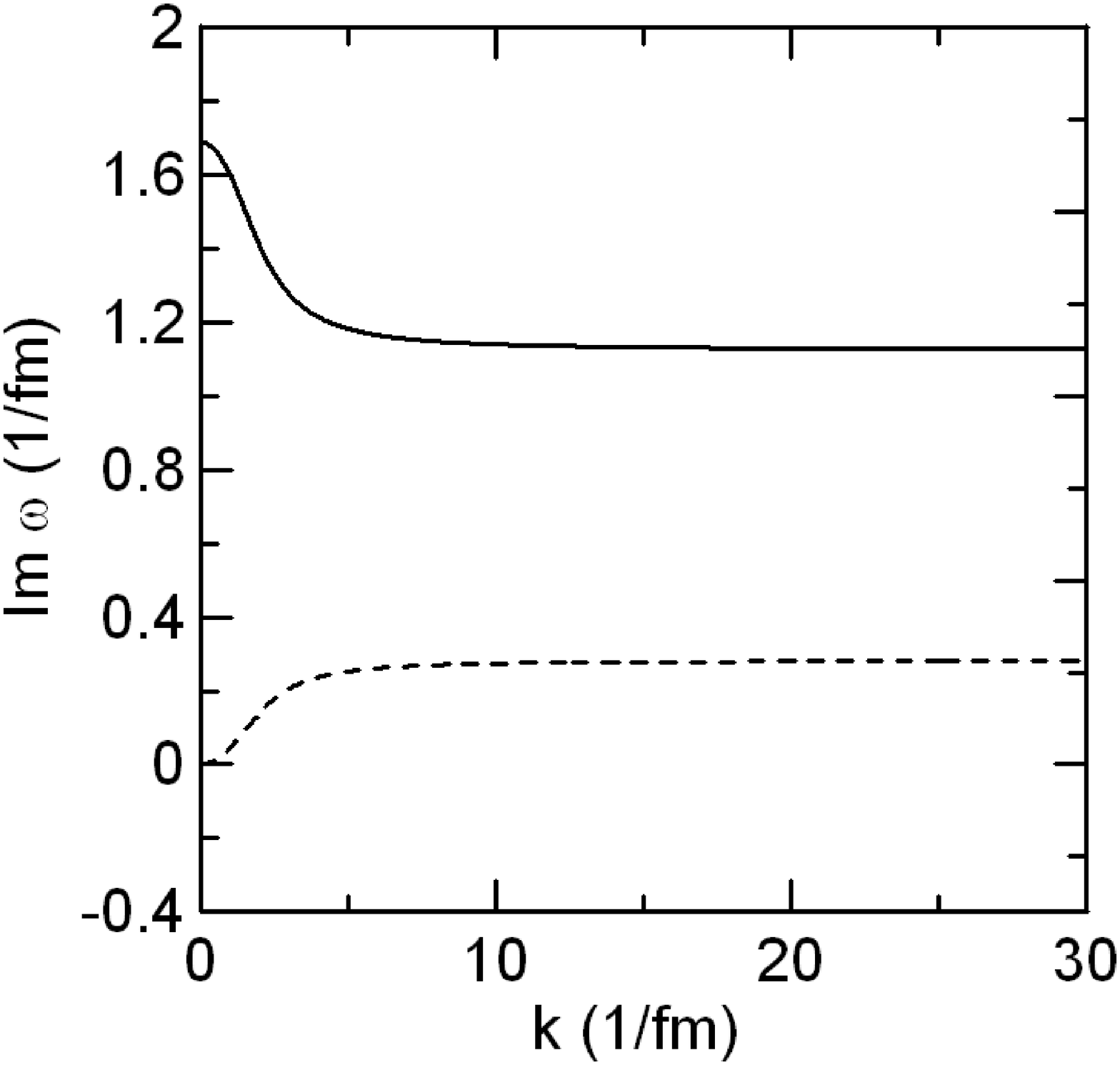}
\end{minipage}
\caption{The real (left panel) and imaginary (right panel) parts of the
frequency in the causal dissipative hydrodynamics at the rest frame for $%
a=0.1$ and $b=6$. There are three modes. One is non-propagating mode (the
solid line), and the other two are propagating modes (the dotted lines). The
two imaginary parts of the propagating modes are degenerated.}
\label{b=6_0}
\end{figure}

In Fig. \ref{b=6_0}, we show the real and imaginary parts of the frequency $%
\omega $ as functions of momentum $k$ for $a=0.1,b=6,$ and the temperature $%
T=200\ MeV$. From the left panel, one can see that the group velocity of the
two propagating modes $\partial Re~\omega /\partial k$ are still slower than
the speed of light. As for the non-propagating mode, we find, from the right
panel, that the imaginary part becomes a constant for the large k, always
remaining positive. This is true for the two propagating modes. The
positivity of the imaginary part guarantees the stability of the hydrostatic
state for plane-wave perturbations. That is, the CD hydrodynamics, with this
parameter set, is causal and stable.

\begin{figure}[tbp]
\begin{minipage}{.45\linewidth}
\includegraphics[scale=0.3]{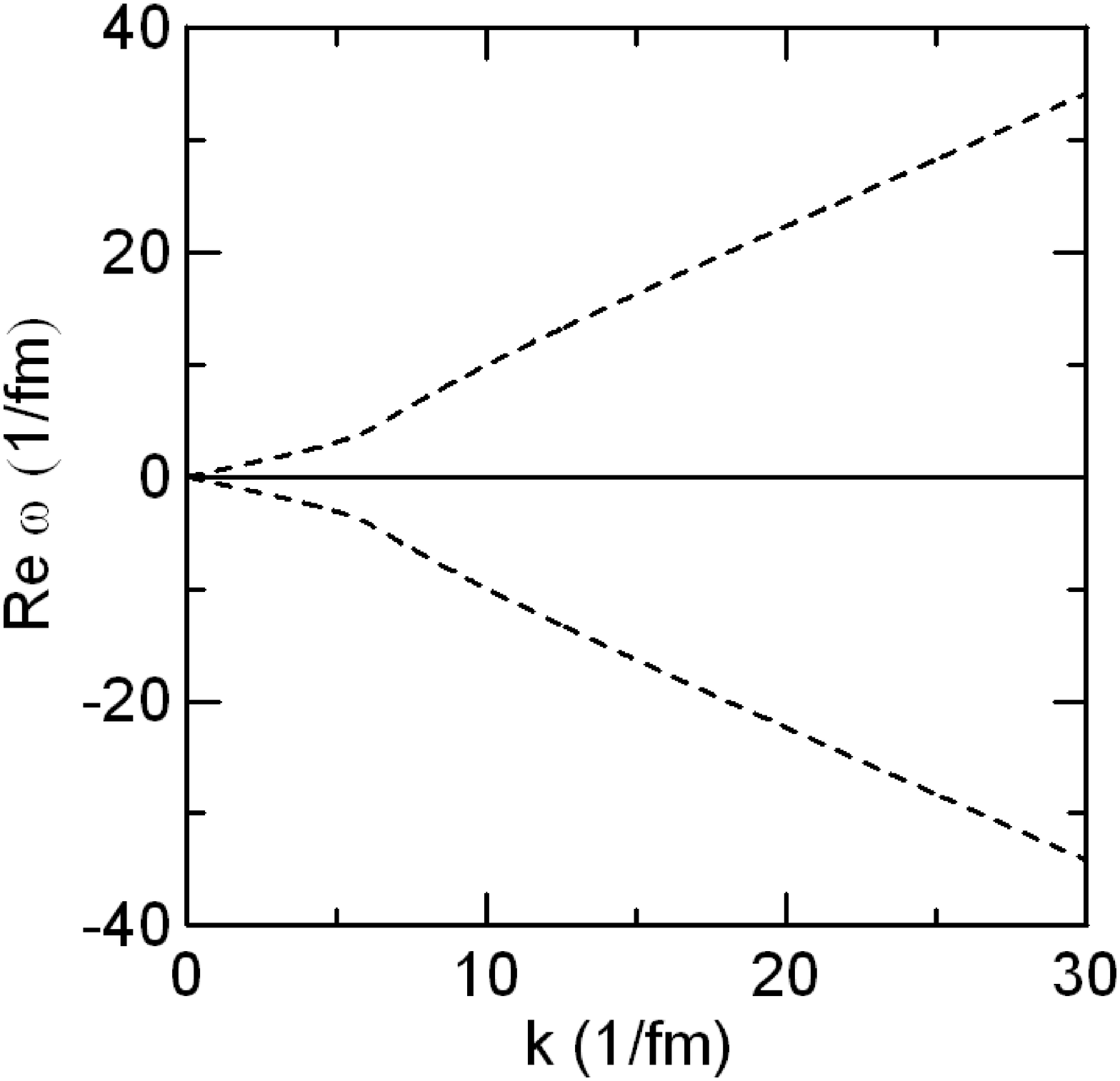}
\end{minipage}
\begin{minipage}{.45\linewidth}
\includegraphics[scale=0.3]{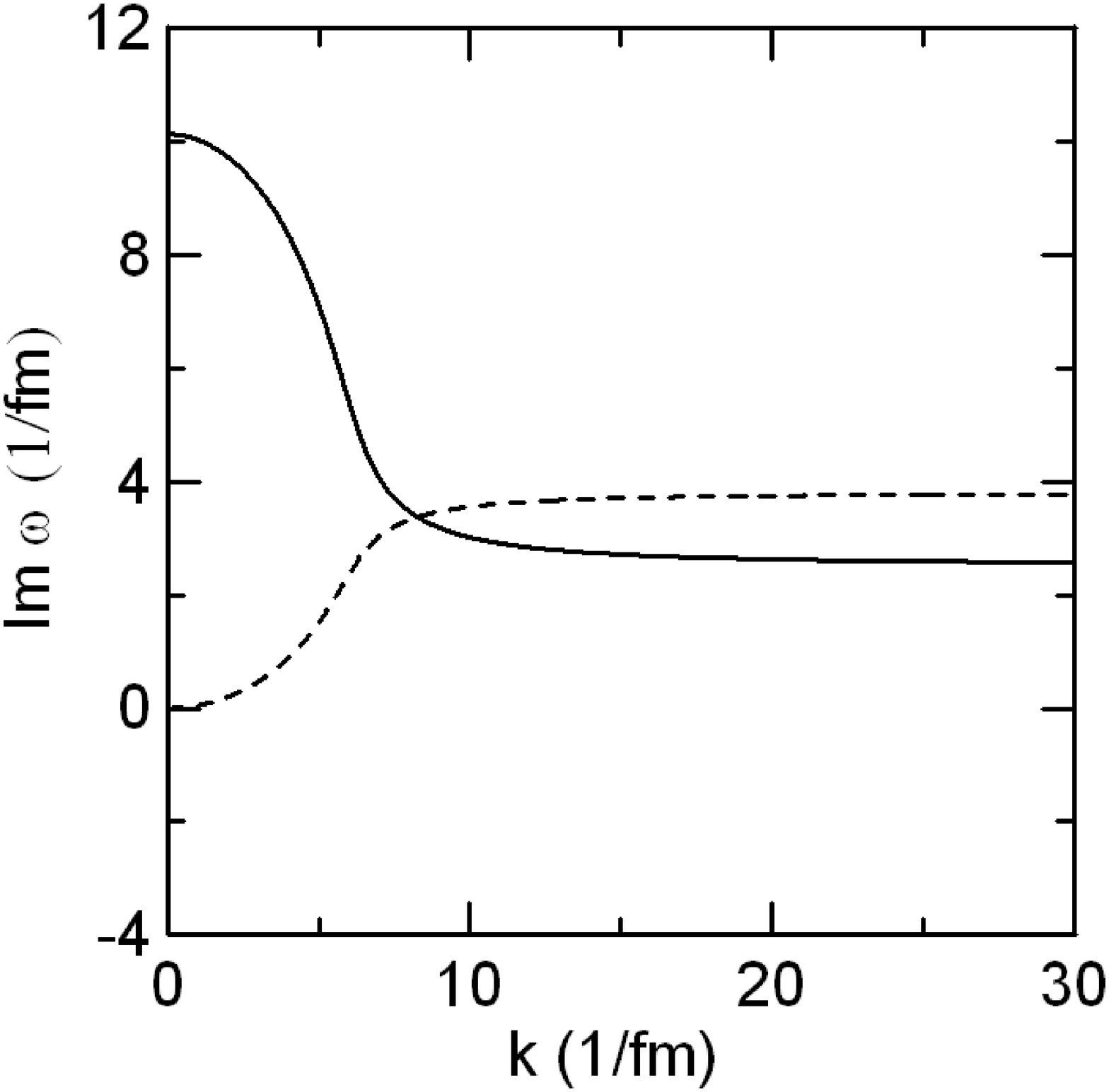}
\end{minipage}
\caption{The real (left panel) and imaginary (right panel) parts of the
frequency in the causal dissipative hydrodynamics at the rest frame for $%
a=0.1$ and $b=1$. There are three modes. One is non-propagating mode (the
solid line), and the other two are propagating modes (the dotted lines). The
two imaginary parts of the propagating modes are degenerated.}
\label{b=1_0}
\end{figure}

However, as was mentioned, the propagation speed of the fluid (\ref%
{eqn:prop_cd}) is affected by the parameter $b$. For the ideal equation of
state, $\alpha =1/3$, the propagation speed exceeds the speed of light if we
use the parameter $b<3/2$. In Fig. \ref{b=1_0}, the real and imaginary parts
of $\omega $ as functions of momentum k are plotted for $a=0.1,b=1$, and the
temperature $T=200MeV$. In this acausal parameter set, $v_{c}$ is larger
than one, and hence, as one can see from the left panel, the propagation
speed of the propagating modes $\partial Re~\omega /\partial k$ exceeds the
speed of light. That is, \textit{even the CD hydrodynamics can be acausal
depending on parameter sets. } However, all the modes have negative
imaginary parts and the theory is still stable.

From these results, one may consider that the problem of acausality is
independent of that of instability. However, as we will see in the next
section, both problems are correlated in relativistic systems.

In this section, we discussed the propagation speed under the linear
approximation. It should be noted that the propagation speed is changed when
the non-linear effect is taken into account. See Appendix \ref{app:nonlinear}
for details.

\section{Stability in general equilibrium frame}

\label{chap:3}

In the previous section, we discussed the stability of a small
perturbation mode around the hydrostatic state. Then we found that even if
the dynamics is not consistent with causality (the LL theory and the CD
hydrodynamics with the acausal parameter set), the hydrostatic states are
still stable. Thus stability is not related with the causality of the theory
in these mode. 
However, these two should be related. Suppose an
acausal propagation of a wave in a covariant theory is allowed. Then an
initial pulse within the light-cone eventually would develop a singular
behavior at the light-cone, since the light-cone cannot be crossed within a
covariant theory. To clarify this point, we will investigate 
the behaviors of the perturbation in a general Lorentz boosted frame.

Let us consider the linear perturbation around the hydrostatic state from
the Lorentz boosted frame moving with the velocity $V$. Then the total
velocity of the fluid is given by 
\begin{eqnarray}
U^{\mu^{\prime}} &=& \gamma_V (\cosh \theta + V \sinh \theta, V \cosh \theta
+ \sinh \theta),  \notag \\
&=& \cosh(\psi + \theta) (1 , \tanh (\psi + \theta)) ,
\end{eqnarray}
where $\tanh \psi = V$.

Substituting this into the energy-momentum tensor and repeating the same
linear analysis around the hydrostatic state with $\psi$ being maintained
constant, the evolution equation for the linear perturbation is given by 
\begin{eqnarray}
A \left( 
\begin{array}{c}
\varepsilon_1 \\ 
\theta_1 \\ 
\Pi_1%
\end{array}
\right) =0,
\end{eqnarray}
where the components of the matrix are 
\begin{eqnarray}
A_{11} &=& (\cosh^2 \psi + \sinh^2 \psi \alpha) (i\omega) + \cosh \psi \sinh
\psi (1+\alpha) (-ik), \\
A_{12} &=& 2(\varepsilon_0 + P_0) \cosh \psi \sinh \psi (i\omega) + w_0
(\cosh^2 \psi + \sinh^2 \psi )(-ik), \\
A_{13} &=& \sinh^2 \psi (i\omega) + \cosh \psi \sinh \psi (-ik), \\
A_{21} &=& \cosh \psi \sinh \psi (1+\alpha) (i\omega) + (\sinh^2 \psi +
\cosh^2 \psi \alpha) (-ik), \\
A_{22} &=& w_0 (\cosh^2 \psi + \sinh^2 \psi ) (i\omega) + 2w_0 \cosh \psi
\sinh \psi (-ik), \\
A_{23} &=& \cosh \psi \sinh \psi (i\omega) + \cosh^2 \psi (-ik), \\
A_{31} &=& 0 , \\
A_{32} &=& \zeta_0 (\sinh \psi (i\omega) + \cosh \psi (-ik) ), \\
A_{33} &=& \tau_{R0}\cosh \psi (i\omega) + \tau_{R0}\sinh \psi (-ik) + 1 .
\end{eqnarray}

Then the dispersion relation is obtained by solving the following equation, 
\begin{eqnarray}
iA \omega^3 + iB k \omega^2 + C \omega^2 + iD k^2 \omega + E k \omega + iF
k^3 + G k^2 = 0,  \label{eqn:moving_omega}
\end{eqnarray}
where 
\begin{eqnarray}
A &=& \cosh \theta ( -1 - (1-v^2_c) \sinh^2 \psi ) , \\
B &=& \sinh \psi ( 1 - (1-v^2_c) + 3(1-v^2_c)\cosh^2 \psi ), \\
C &=& -\frac{1}{\tau_{R0}} ( \alpha + (1-\alpha)\cosh^2 \psi ), \\
D &=& \cosh \psi (-3 (1-v^2_c) \cosh^2 \psi + 2(1- v^2_c) + 1 ), \\
E &=& \frac{2}{\tau_{R0}}(1-\alpha) \cosh \psi \sinh \psi , \\
F &=& \sinh \psi (- 1 + (1-v^2_c)\cosh^2 \psi ), \\
G &=& \frac{1}{\tau_{R0}} (1 - (1-\alpha)\cosh^2 \psi).
\end{eqnarray}
One can easily see that the behavior of the frequency $\omega$ depends on
the choice of $v_c$, which is the propagation speed of the fluid defined by
Eq. (\ref{eqn:prop_cd}).

\subsection{Landau-Lifshitz theory}

As we discussed, the Landau-Lifshitz theory has two modes in the local rest
frame. In the Lorentz boosted frame, however, we have three modes, because
of the following reason. In this case, the coefficients of Eq. (\ref%
{eqn:moving_omega}) are given by; 
\begin{eqnarray}
A &=& \zeta_0 \cosh \psi \sinh^2 \psi, \\
B &=& \zeta_0 \sinh \psi (1-3 \cosh^2 \psi), \\
C &=& -(\varepsilon_0 + P_0)(\alpha + (1-\alpha)\cosh^2 \psi), \\
D &=& \zeta_0 \cosh \psi (-2 + 3 \cosh^2 \psi), \\
E &=& 2 (\varepsilon_0 + P_0) (1-\alpha) \cosh \psi \sinh \psi, \\
F &=& -\zeta_0 \sinh \psi \cosh^2 \psi, \\
G &=& (\varepsilon_0 + P_0)(1-(1-\alpha)\cosh^2 \psi).
\end{eqnarray}
The coefficient $A$ disappears only in the rest frame ($\psi=0$). That is,
there exists a gap in the calculations of the rest frame and the moving
frame.

\begin{figure}[tbp]
\begin{minipage}{.45\linewidth}
\includegraphics[scale=0.3]{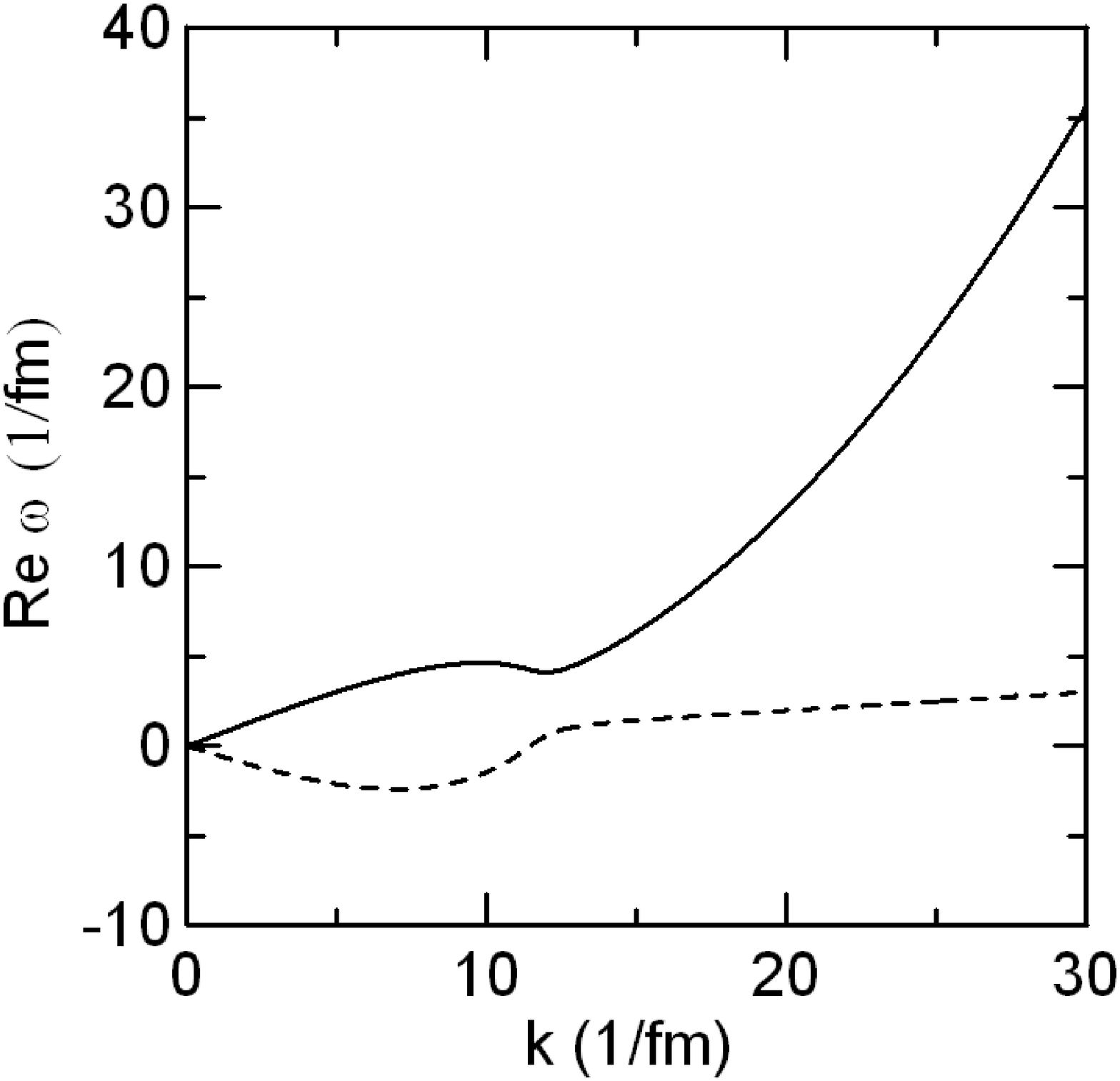}
\end{minipage}
\begin{minipage}{.45\linewidth}
\includegraphics[scale=0.3]{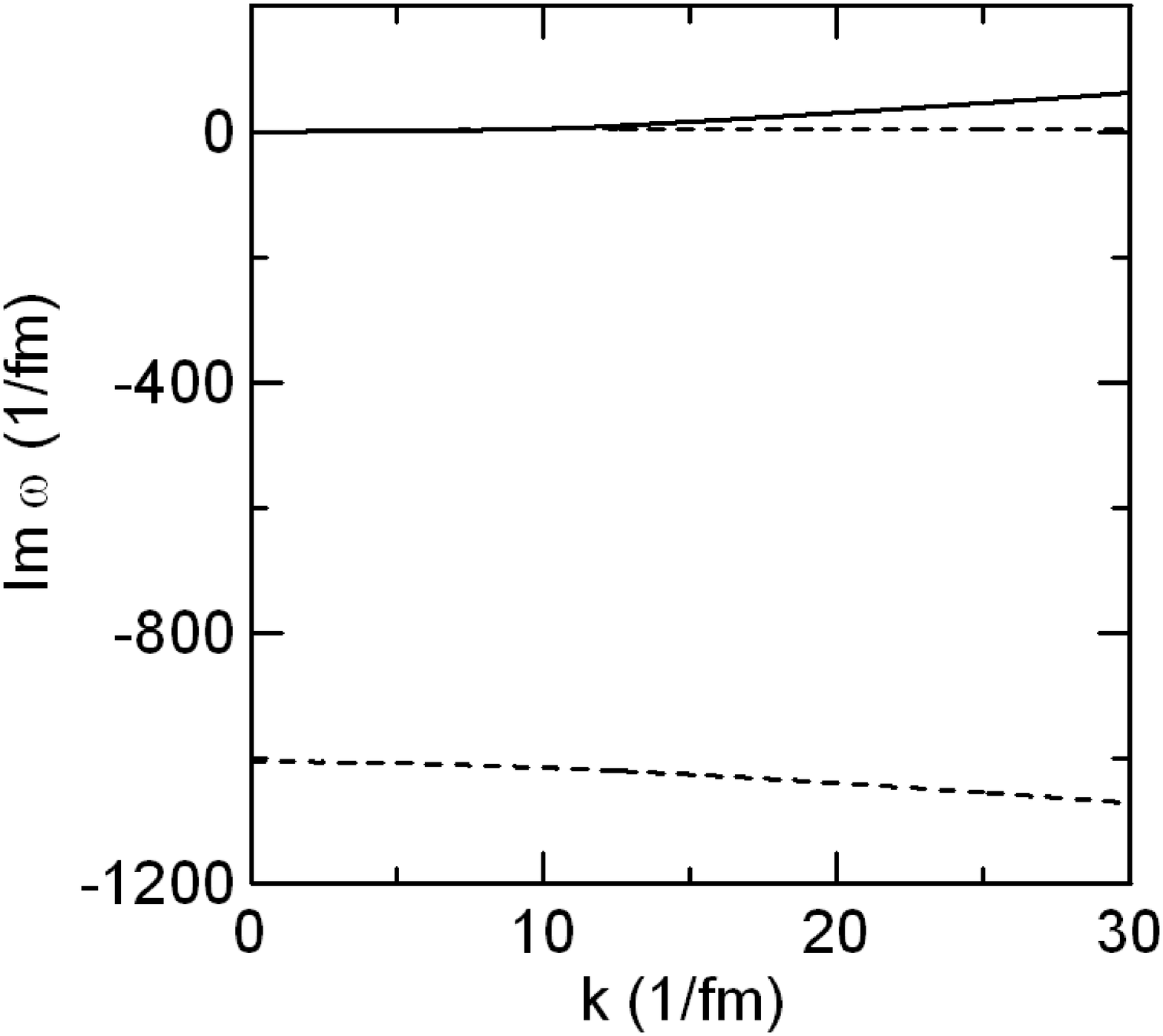}
\end{minipage}
\caption{The real (left panel) and imaginary (right panel) parts of the
frequency in the Landau-Lifshitz theory in the Lorentz boosted frame with
the velocity $v=0.1$. There are three propagating modes. One of the modes is
denoted by the solid line. The remaining two modes are degenerated and
plotted by the dotted line. One of the imaginary parts (one of the dotted
lines) is negative. }
\label{LL_0_1}
\end{figure}

In Fig. \ref{LL_0_1}, the real and imaginary parts of the frequency $\omega$
are shown for $a=0.1$ and $T=200$ MeV at $V = 0.1$. One can see that all the
three modes are propagating modes. The real parts of the two propagating
modes denoted by the dotted line are degenerated, but the imaginary part of
one of them is negative. Thus the LL theory is unstable in the Lorentz
boosted frame.

\subsection{Causal dissipative hydrodynamics}

\begin{figure}[tbp]
\begin{minipage}{.45\linewidth}
\includegraphics[scale=0.3]{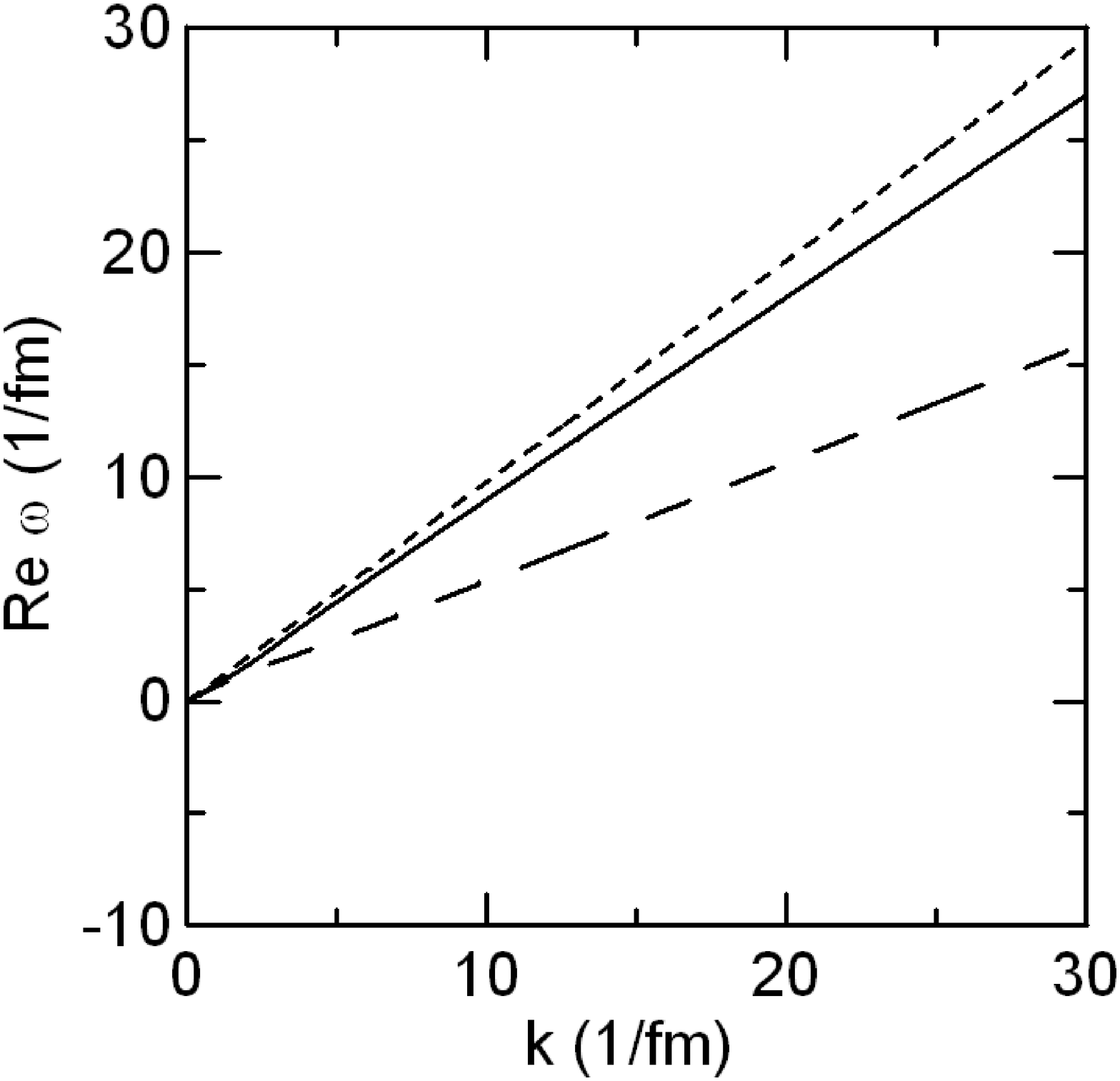}
\end{minipage}
\begin{minipage}{.45\linewidth}
\includegraphics[scale=0.3]{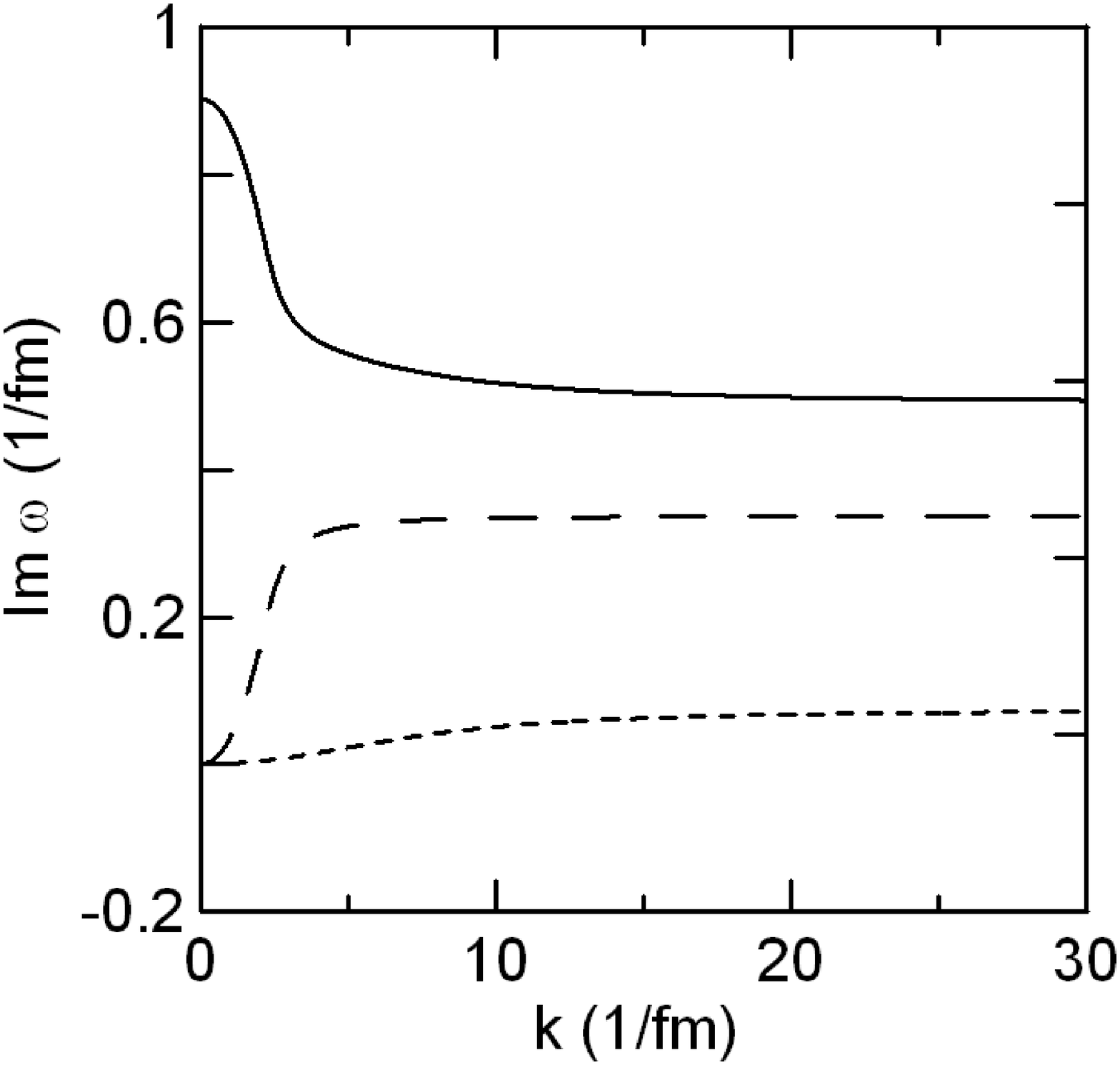}
\end{minipage}
\caption{The real ( left panel) and imaginary ( right panel) parts of the
frequency in the causal dissipative hydrodynamics in the Lorentz boosted
frame with the velocity $v=0.9$ for $a=0.1$ and $b=1$. There are three modes
denoted by the solid, dotted and dashed lines, respectively. All modes have
positive imaginary parts.}
\label{b=6_v=0_9}
\end{figure}

First, we consider the causal parameter set, $a=0.1$ and $b=6$ where the
propagation speed (\ref{eqn:prop_cd}) is slower than the speed of light. In
Fig. \ref{b=6_v=0_9}, the real and imaginary parts of the frequency $\omega$
are shown for $T=200$ MeV at $V=0.9$. From the behaviors of the real parts,
one can see that the three group velocities become close to the speed of
light, but never exceed. On the other hand, all three propagating modes have
positive imaginary parts. As far as we checked, this theory does not becomes
acausal and does not have a negative imaginary part for any $V$. Thus in
this causal parameter set, the CD hydrodynamics is consistent with causality
and stable even in the Lorentz boosted frame.

\begin{figure}[tbp]
\begin{minipage}{.45\linewidth}
\includegraphics[scale=0.3]{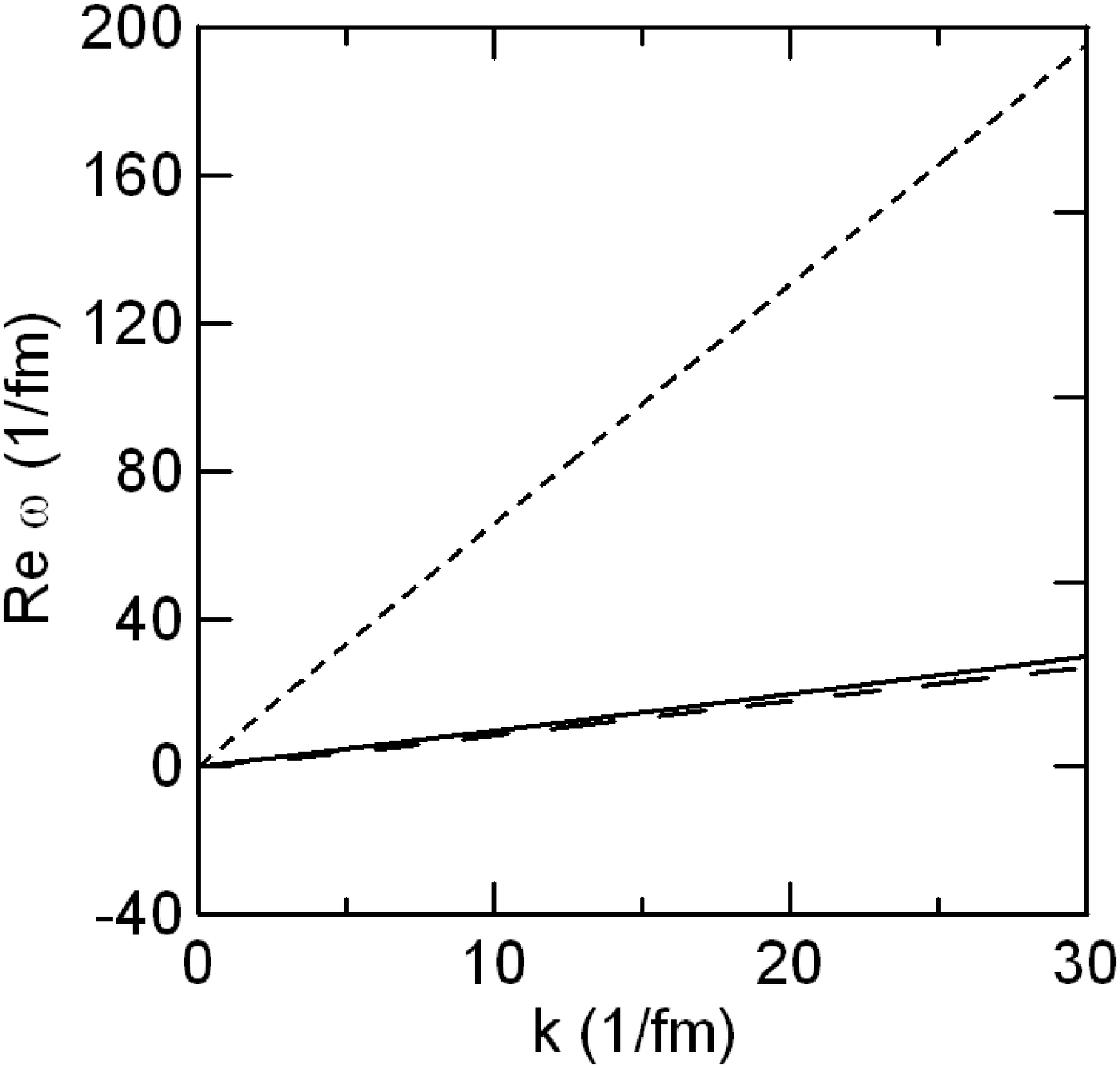}
\end{minipage}
\begin{minipage}{.45\linewidth}
\includegraphics[scale=0.3]{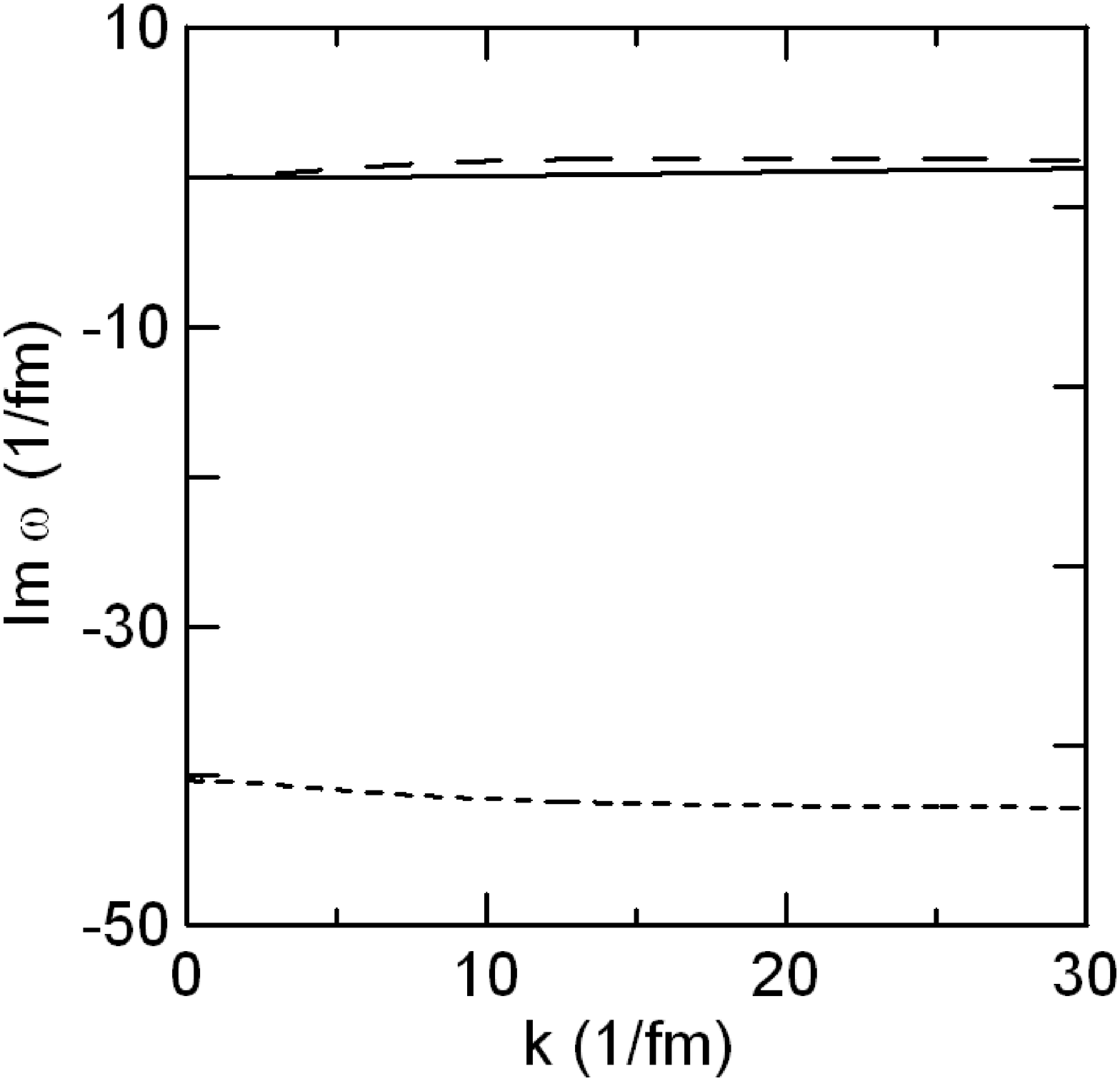}
\end{minipage}
\caption{The real (left panel) and imaginary (right panel) parts of the
frequency in the causal dissipative hydrodynamics in the Lorentz boosted
frame with the velocity $v=0.9$ for $a=1$ and $b=1$. There are three modes
denoted by the solid, dotted and dashed lines, respectively. One mode (the
dotted line) has a negative imaginary part.}
\label{b=1_v=0_9}
\end{figure}

However, this is not true for the acausal parameter set, for example, $a=0.1$
and $b=1$. In Fig. \ref{b=1_v=0_9}, we show the real and imaginary parts of
the frequency $\omega$. There are three propagating modes denoted by the
solid, dashed and dotted line. It is clear that the group velocity is faster
than the speed of light. Interestingly enough, one of the propagating modes
denoted by the dotted line has a negative imaginary part. Thus the CD
hydrodynamics with acausal parameter set is unstable in the Lorentz boosted
frame.

This result means that causality and stability are correlated and
instability is induced by acausality in the relativistic dissipative
hydrodynamics. A consistent theory should not change its stability depending
on the choice of the frames. Thus, the LL theory is not consistent theories
as candidates for the relativistic dissipative hydrodynamics. 
This is so also for the CD hydrodynamics with acausal parameter sets.

We should stress, as a matter of fact, that we cannot implement
stable numerical calculations of the CD hydrodynamics when we use acausal
parameter sets.

The stability of the LL theory from a general frame is discussed also in 
\cite{his2} in a different context. See Appendix \ref{app:his} for details.

\section{Stability around Scaling Solution}

\label{chap:4}

In this section, we discuss the stability around the scaling solution. The
scaling variables $\tau $ and $y$ are defined by 
\begin{eqnarray}
\tau =\sqrt{t^{2}-z^{2}},~~~~~y=\frac{1}{2}\ln \left[ \frac{t+z}{t-z}\right]
.
\end{eqnarray}

By using these variables, the equations for the conservation of
energy-momentum are reexpressed as 
\begin{eqnarray}
(\tau \partial _{\tau }+\tanh (\theta -y\pm \theta _{s})\partial _{y})(\phi
\pm \theta )-\frac{\tau }{\cosh (\theta -y)\pm c_{s}\sinh (\theta -y)}R^{-1}%
\left[ c_{s}\nabla \theta \pm \left( D\theta +\frac{1}{\Pi }\nabla \Pi
\right) \right] =0, \label{eqn:sc_ori1}
\end{eqnarray}%
where $\tanh \theta _{s}=\sqrt{\alpha }$. Here we define the Reynolds number 
$R^{-1}=-\Pi /(Ts)$, which reproduces the definition of \cite{kouno} in the
vanishing relaxation time limit. The new variable $\phi $ satisfies the
following relation, 
\begin{eqnarray}
d\phi =\sqrt{\alpha }d\ln s=\frac{1}{\sqrt{\alpha }}d\ln T.
\end{eqnarray}%
The equation for the viscosity is 
\begin{eqnarray}
\tau _{R}\partial _{\tau }\Pi +\Pi =-\zeta \left( \sinh (\theta -y)\partial
_{\tau }+\cosh (\theta -y)\frac{1}{\tau }\partial _{y}\right) \theta . \label{eqn:sc_ori2}
\end{eqnarray}

We assume that the velocity of the fluid is given by 
\begin{eqnarray}
\tanh \theta =\frac{z}{t}=\tanh y.
\end{eqnarray}%
This scaling ansatz is considered to be valid near the central rapidity
region. Thus, the equations of the scaling solution are given by setting $%
\theta =y$, \ 
\begin{eqnarray}
&&\tau \partial _{\tau }\phi _{0}+(1-R_{0}^{-1})c_{s}^{0}=0, \\
&&\tau _{R0}\partial _{\tau }\Pi _{0}+\Pi _{0}=-\frac{\zeta _{0}}{\tau }.
\end{eqnarray}%
In the LL limit ($\tau _{R}\rightarrow 0$), the equations are reduced to
those obtained in \cite{kouno}.

To see the stability of the scaling solution, we consider the linear
perturbation as follows, 
\begin{eqnarray}
\theta &=& y + \delta \theta, \\
\phi &=& \phi_0 + \delta \phi.
\end{eqnarray}
Substituting them into Eqs. (\ref{eqn:sc_ori1}) and (\ref{eqn:sc_ori2}), the
evolution equations of the linear perturbations are given by 
\begin{eqnarray}
&& \tau \partial_{\tau}\delta \phi + (1-R^{-1}_0)c^0_s \partial_y \delta
\theta + (1-R^{-1}_0) \left( \frac{\partial c_s}{\partial \phi} \right)_0
\delta \phi - \delta R^{-1} c^0_s = 0 ,  \label{eqn:lphi} \\
&& \tau (1-R^{-1}_0) \partial_{\tau}\delta \theta + c^0_s \partial_y \delta
\phi + (1-(c^0_s)^2) \delta \theta -(1-(c^0_s)^2) \delta \theta R^{-1}_0 - 
\frac{R^{-1}_0}{\Pi_0} \partial_y \delta \Pi + \frac{1}{\Pi_0}\frac{R^{-1}_0%
}{\tau_R}(\tau \Pi_0 + \zeta_0 ) \delta \theta = 0,  \label{eqn:ltheta} \\
&& \tau_{R0} \partial_{\tau}\delta \Pi + \delta \Pi = \left( \frac{\partial
\ln \tau_{R}}{\partial \phi} \right)_0 (\Pi_0 + \frac{\zeta_0}{\tau})\delta
\phi -\frac{\zeta_0}{\tau} \left(\frac{\partial \ln \zeta}{\partial \phi}
\right)_0 \delta \phi - \zeta_0 \frac{\partial_y\delta \theta}{\tau},
\label{eqn:lpi}
\end{eqnarray}
where 
\begin{eqnarray}
\delta R^{-1} = - \frac{1}{(Ts)_0} \delta \Pi + \frac{\Pi_0}{(Ts)_0}\left( 
\frac{\partial \ln(Ts)}{\partial \phi} \right)_0 \delta \phi,
\end{eqnarray}
with the scaling solutions $\phi_0$ and $\Pi_0$.

By using the Fourier transform for $y$, 
\begin{eqnarray}
A(\tau,y) = \int dk e^{-iky}A(\tau,k),
\end{eqnarray}
Eqs. (\ref{eqn:lphi}), (\ref{eqn:ltheta}) and (\ref{eqn:lpi}) are summarized
as 
\begin{eqnarray}
\tau\partial_{\tau}X = A X,  \label{eqn:eqX}
\end{eqnarray}
where 
\begin{eqnarray}
X = \left( 
\begin{array}{c}
\delta \phi \\ 
\delta \theta \\ 
\delta \ln \Pi%
\end{array}
\right),  \label{eqn:X}
\end{eqnarray}
and 
\begin{eqnarray}
A = \left( 
\begin{array}{ccc}
- (1-R^{-1}_0) \left( \frac{\partial \sqrt{\alpha}}{\partial \phi} \right)_0
- R^{-1}_0 \left( 1 + \alpha \right) & - (1-R^{-1}_0) \sqrt{\alpha} (-ik) & 
\sqrt{\alpha} R^{-1}_0 \\ 
- \sqrt{\alpha} \frac{(-ik)}{(1-R^{-1}_0)} & - (1-\alpha) - \frac{(R^{-1}_0 
\hat{\tau} - \frac{1}{b} )}{(1-R^{-1}_0)} & R^{-1}_0 \frac{(-ik)}{%
(1-R^{-1}_0)} \\ 
- \sqrt{\alpha} \left( \hat{\tau} - \frac{1}{R^{-1}_0 b}\right) + \frac{1}{%
\sqrt{\alpha}R^{-1}_0 b} & \frac{-ik}{R^{-1}_0 b} & -\frac{1}{R^{-1}_0 b}%
\end{array}
\right),
\end{eqnarray}
where $\hat{\tau}=\tau/\tau_{R0}$ and 
\begin{eqnarray}
\frac{\partial \ln (Ts)}{\partial \phi} &=& \sqrt{\alpha} + \frac{1}{\sqrt{%
\alpha}}, \\
\frac{\partial \ln \tau_{R0}}{\partial \phi} &=& - \sqrt{\alpha}, \\
\frac{\partial \ln \zeta}{\partial \phi} &=& \frac{1}{\sqrt{\alpha}}.
\end{eqnarray}
Note that we consider the ideal gas equation of state where $\alpha = 1/3$
and $\left( \partial \sqrt{\alpha}/\partial \phi \right)_0 =0$.

One can see that the scaling solution is \textit{unstable} when $\hat{\tau}%
>1/b$, for $k=0$ and $R_{0}^{-1}=1+\epsilon $, and $\hat{\tau}<1/b$, for $%
k=0 $ and $R_{0}^{-1}=1-\epsilon $, because the equation for $\delta \theta $
becomes decoupled and $A_{22}$ becomes positive. Here $\epsilon $ is a small
arbitrary constant.

In general, the analysis of the stability is non-trivial unless the matrix $A
$ can be diagonalized. We discuss the stability in the Lyapunov direct
method used in \cite{kouno}. In this approach, the stability of the theory
is analyzed by introducing the Lyapunov function, which characterizes the
deviation from the scaling solution. The Lyapunov function should be 1)
positive definite and 2) monotonically decreasing function with
respect to the measure of the distance of the perturbed solution from the
non-perturbed one. If we can find the Lyapunov function at given $k$, $b$, $%
\hat{\tau}$ and $R_{0}$, the scaling solution is stable for the parameter
set. If the assumed Lyapunov function is found to be a monotonically
increasing function then the scaling solution is unstable for the respective
parameter set. See Appendix \ref{app:sta} for details.

\begin{figure}[tbp]
\begin{minipage}{.45\linewidth}
\includegraphics[scale=0.3]{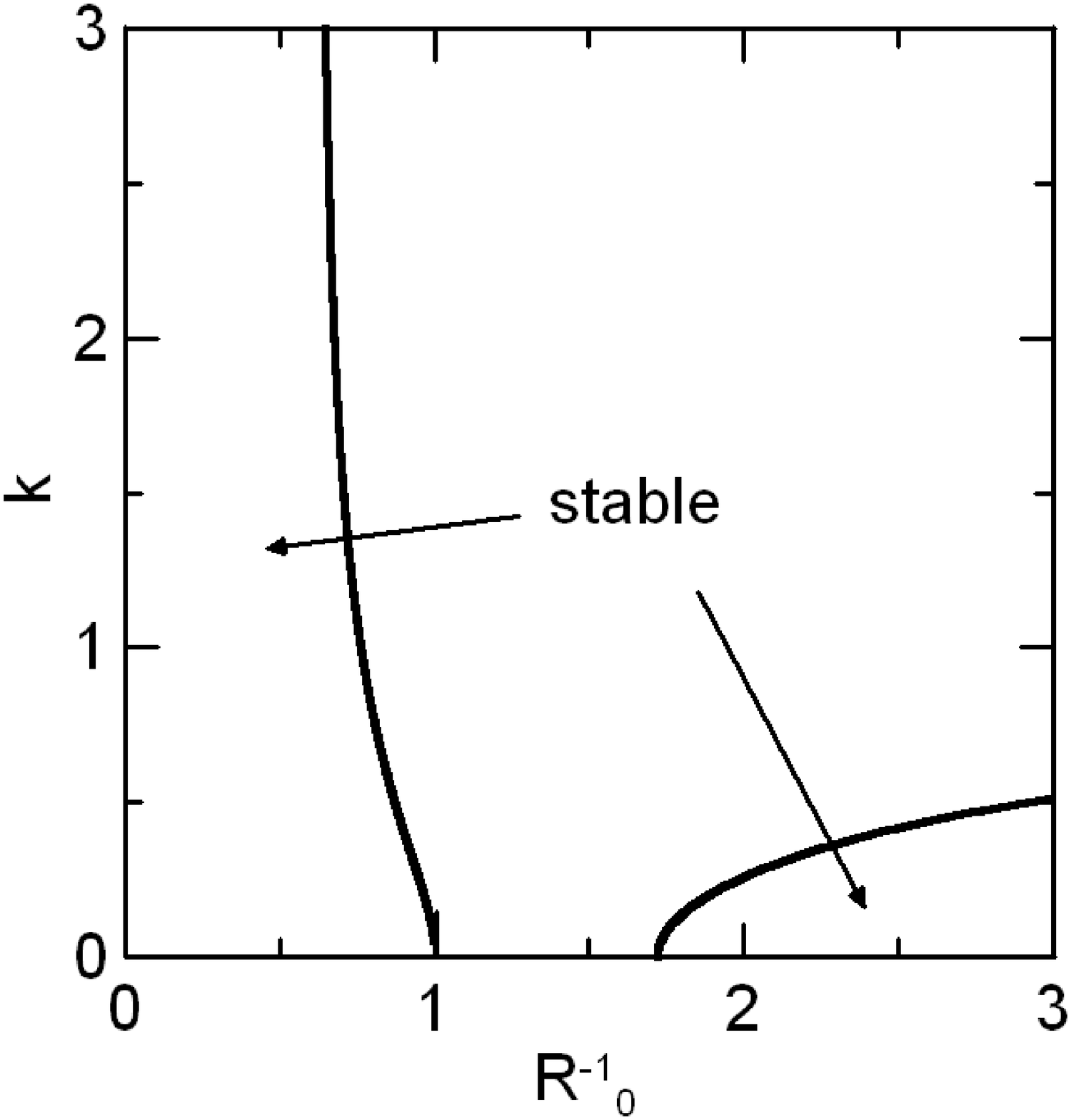}
\caption{The phase diagram of the stability in the LL theory as a function of $k$ and $R_0$.
There are two stable regions. 
}
\label{phase_LL}
\end{minipage}
\hspace{1cm} 
\begin{minipage}{.45\linewidth}
\includegraphics[scale=0.3]{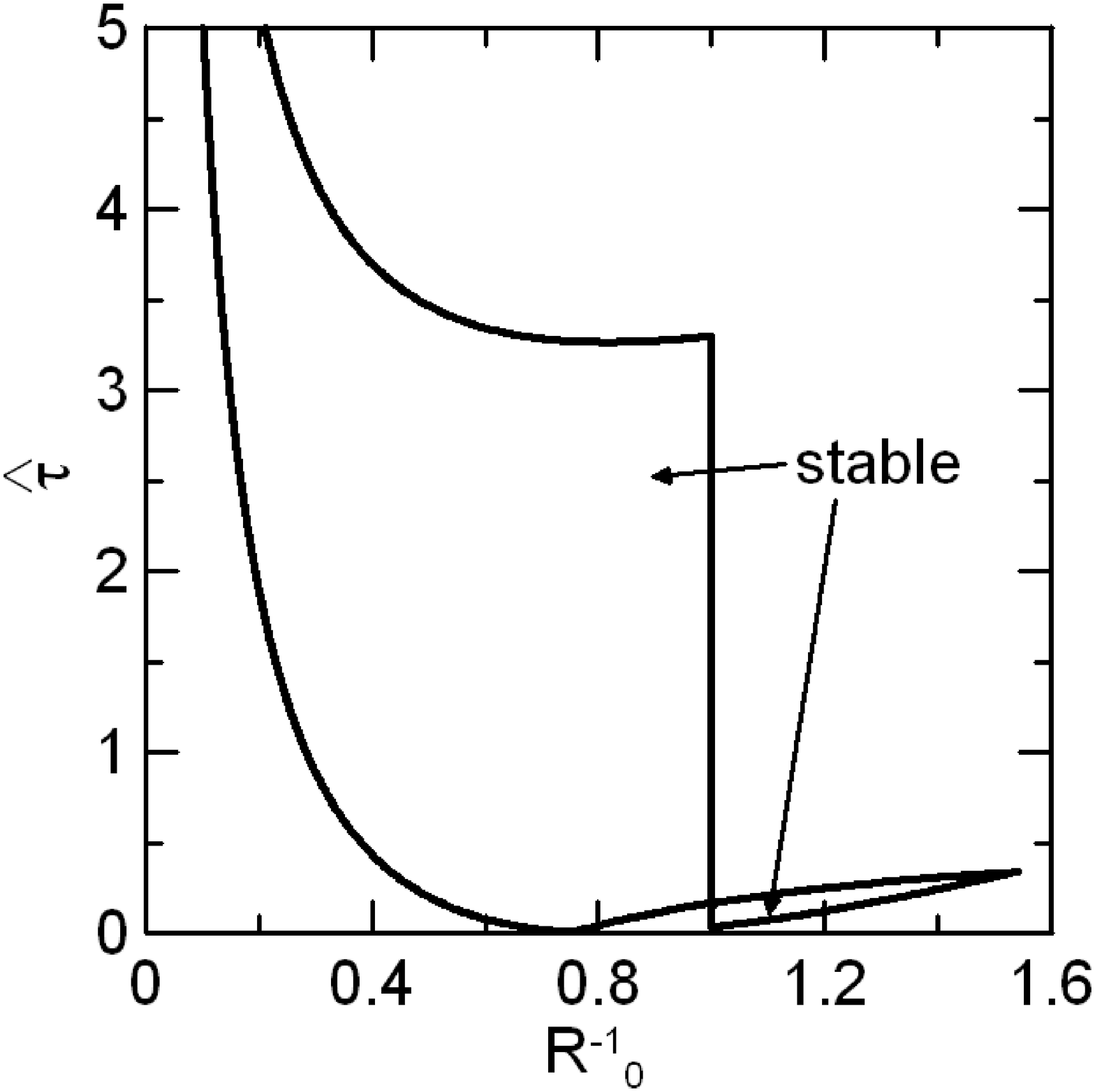}
\caption{The phase diagram of the stability in the CD hydrodynamics as a function 
of $\hat{\tau}$ and $R_0$. We set $k=0$ and $b=6$. 
There are two stable regions. Even though $R_0 \ge 1$, 
the scaling solution of the CD hydrodynamics can be unstable.}
\label{phase_IS_k=0_b=6}
\end{minipage}
\end{figure}

However, we should be noted that the stable region predicted in the Lyapunov
direct method is normally under- estimated, unless we exhaust every
possible Lyapunov functions. As an example, let us consider the limit of
the LL theory which is realized when we set $\tau _{R}=\partial \ln \tau
_{R}/\partial \phi =0$, in Eqs. (\ref{eqn:lphi}), (\ref{eqn:ltheta}) and (%
\ref{eqn:lpi}). In Fig. \ref{phase_LL}, we show the phase diagram for the
stability, on the $k-R_{0}^{-1}$ plane, assuming the Lyapunov function $%
V=|\delta \phi |^{2}+|\delta \theta |^{2}$. There are two stable regions;
one is in $R_{0}\geq 1$ and the other in the small $k$ and small $R_{0}$. We
found that the scaling solution is stable also on the line of $R_{0}=1$ by
solving the same $V$, but it is not shown in the phase diagram for
simplicity. To confirm more precisely the stable regions, we have to study
various possible Lyapunov functions. Then the real stable regions, in
general, can distribute in broader regions on the phase diagram. As a matter
of fact, the phase diagram obtained by Kouno et al. shows that the LL theory
is stable for any $k$ in all region of $R_{0}\geq 1$ \cite{kouno}. On the
other hand, we could not find unstable regions for this Lyapunov function.
The stability of the domain between the two stable regions is not confirmed.
However, from the analysis of Kouno et al., a part of the unconfirmed region
should be an unstable region \cite{kouno}.

To analize the CD hydrodynamics, we first consider the following Lyapunov
function $V$, 
\begin{eqnarray}
V=X^{\dagger }X=(\delta \phi )^{2}+(\delta \theta )^{2}+(\delta \ln \Pi
)^{2}.
\end{eqnarray}%
Note that the time evolution of $V$ is given by, 
\begin{eqnarray}
\tau \partial _{\tau }V=X^{\dagger }(A^{\dagger }+A)X.
\end{eqnarray}

For $R_{0}=1$, we can eliminate the variable $\delta \phi $ from the
equations. In this case, we will consider the simpler Lyapunov function, 
\begin{eqnarray}
V=Y^{\dagger }Y=(\delta \phi )^{2}+(\delta \ln \Pi )^{2}.
\end{eqnarray}%
Here the evolution equation of $Y$ is given by 
\begin{eqnarray}
\tau \partial _{\tau }Y=BY,
\end{eqnarray}%
where 
\begin{eqnarray}
Y=\left( 
\begin{array}{c}
\delta \phi \\ 
\delta \ln \Pi%
\end{array}%
\right) ,
\end{eqnarray}%
and 
\begin{eqnarray}
B=\left( 
\begin{array}{cc}
-\left( 1+\alpha \right) & \sqrt{\alpha } \\ 
-\sqrt{\alpha }\left( \hat{\tau}-\frac{1}{b}\right) +\frac{1}{\sqrt{\alpha }b%
}+\frac{\sqrt{\alpha }k^{2}}{b\hat{\tau}-1} & -\frac{1}{b}-\frac{k^{2}}{b%
\hat{\tau}-1}%
\end{array}%
\right) .
\end{eqnarray}%
In this case, we should discuss the eigen values of $B^{\dagger }+B$.

There are three parameters, $k$, $\hat{\tau}$ and $R_{0}$, fixing $b=6$.
First, we discuss the homogeneous perturbation setting $k=0$ and calculate
the phase diagram in the $\hat{\tau}-R_{0}^{-1}$ plane, as is shown in Fig. %
\ref{phase_IS_k=0_b=6}. There are two stable regions; one is very small and
located in $R_{0}<1$, and the other is larger and located in $R_{0}\geq 1$.
This result is consistent with the instability of the scaling solution in $%
\hat{\tau}>1/b$ for $R_{0}^{-1}=1+\epsilon $ and $\hat{\tau}<1/b$ for $%
R_{0}^{-1}=1-\epsilon $. It should, however, be noted that one can see that
the scaling solution is always stable on $R_{0}^{-1}=0$, which can be seen
from the behavior of Eq. (\ref{eqn:eqX}) itself. On the other hand, we could
not find unstable regions. As a matter of fact, we will discuss the
stability by using various Lyapunov functions in the following, but still
cannot find any unstable regions.

As we have pointed out, the Lyapunov direct method normally underestimates
the stable region. To fix the stable region, we have to discuss as many
Lyapunov functions as possible. In this work, we analyzed the phase diagram
with more three different functions: $V^{\prime }=|\delta \ln s|^{2}+|\delta
\theta |^{2}+|\delta \ln \Pi |^{2}$, $V^{\prime \prime }=|\delta \ln
s|^{2}+|\delta \theta |^{2}+|R_{0}^{-1}\delta \ln \Pi |^{2}$ and $V^{\prime
\prime \prime }=|\delta \phi |^{2}+|\delta \theta |^{2}+|R_{0}^{-1}\delta
\ln \Pi |^{2}$. We found that the qualitative structure of the phase diagram
is independent of the choice of these functions. We show the result of $%
V^{\prime \prime \prime }$ in Fig. \ref{phase_IS_k=0_b=6_v2}, for which we
obtained the maximum stable region (See Appendix \ref{app:V} for detailed
form of the equation). This phase diagram shows that most part of the phase
diagram in $R_{0}<1$ still belongs to the unconfirmed region. As for the
region of $R_{0}\geq 1$, we found that the stable region strongly depends on
the choice of the Lyapunov function, and most of the physically accessible
region is confirmed to be stable. As a matter of fact, the dashed lines in
Fig. \ref{phase_IS_k=0_b=6_v2} shows the trajectories of the scaling
solutions for $a=0.1$ (left) and $a=1$ (right), and one can see that most of
the trajectories belongs to the stable region in the phase diagram.

\begin{figure}[tbp]
\begin{minipage}{.45\linewidth}
\includegraphics[scale=0.3]{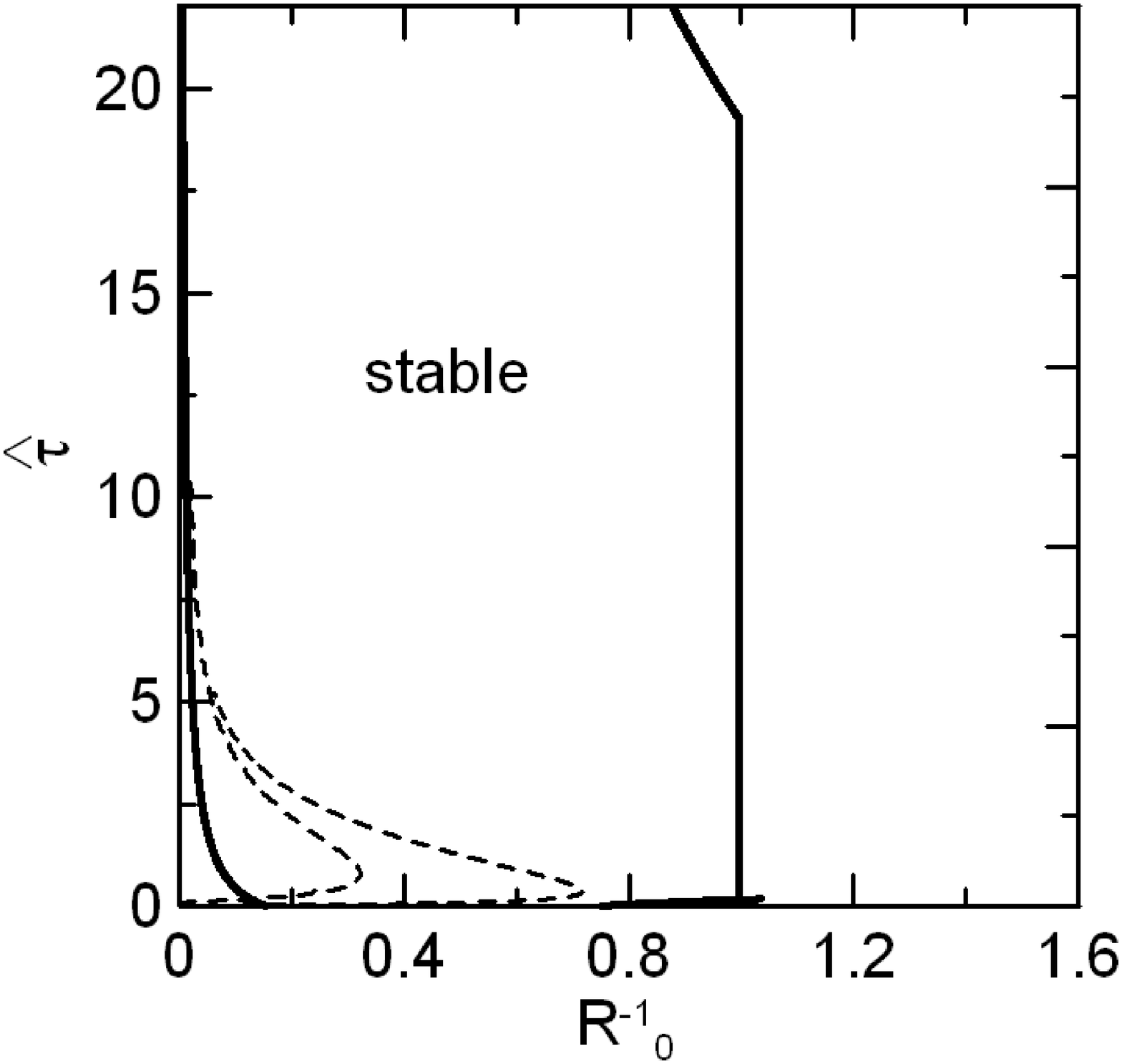}
\caption{The phase diagram of the stability in the CD hydrodynamics which is 
calculated with the Lyapunov function $V'''$. We set $k=0$ and $b=6$.
Compared to Fig. (\ref{phase_IS_k=0_b=6}), the stable region is enlarged.}
\label{phase_IS_k=0_b=6_v2}
\end{minipage}
\hspace{1cm} 
\begin{minipage}{.45\linewidth}
\includegraphics[scale=0.3]{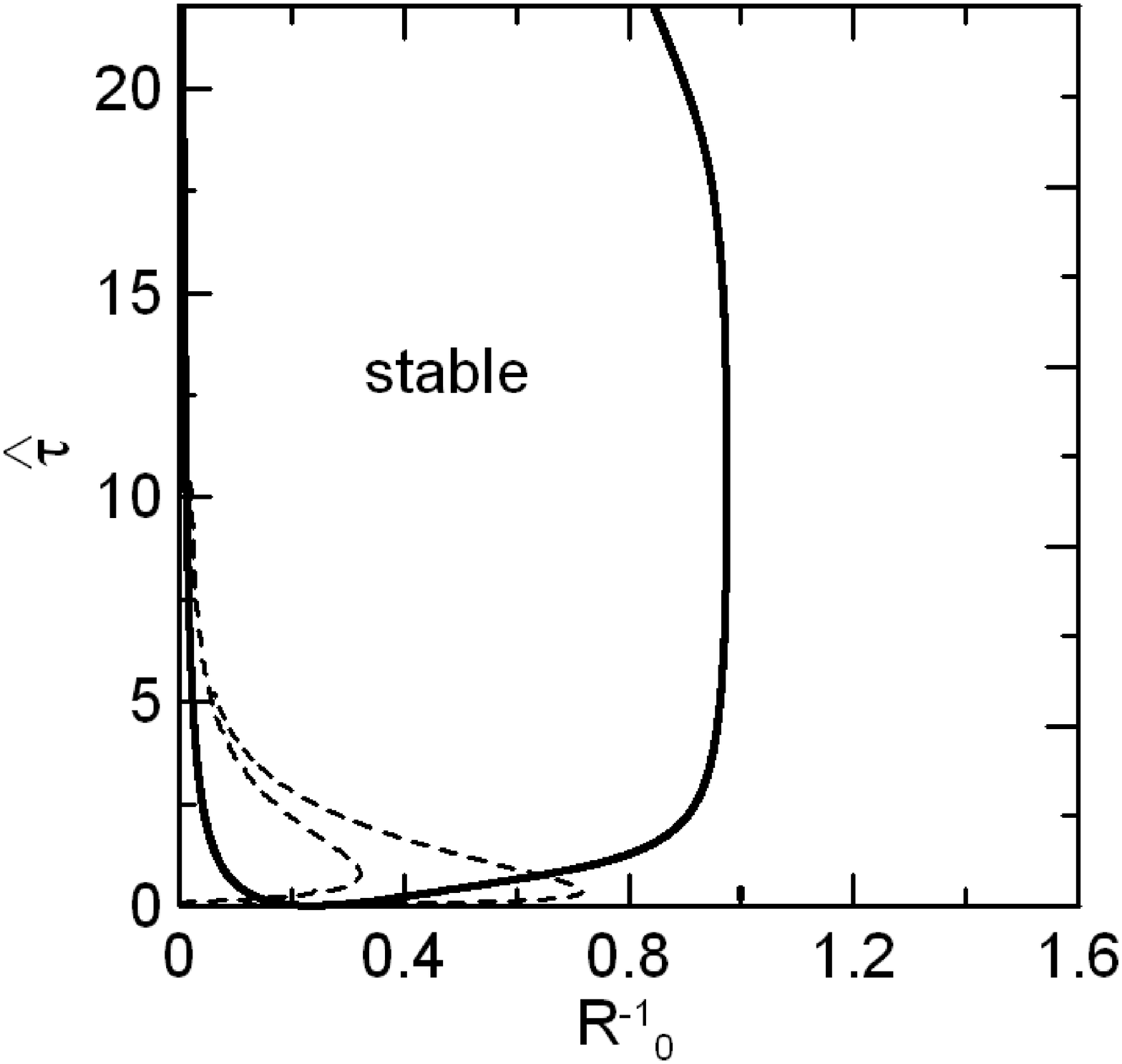}
\caption{The phase diagram of the stability in the CD hydrodynamics which is 
calculated with the Lyapunov function $V'''$. We set $k=1$ and $b=6$. 
The stable region for homogeneous perturbation ($k=0$) at low $\hat{\tau}$ around $R_0 = 1$ 
is changed to the unconfirmed region for the inhomogeneous perturbation ($k=1$).}
\label{phase_IS_k=1_b=6_v2}
\end{minipage}
\end{figure}

In this sense, we conclude that the scaling solution is stable for
homogeneous perturbation. To see the stability for the inhomogeneous
perturbation, we need to discuss the phase diagram for finite $k$. In Fig. %
\ref{phase_IS_k=1_b=6_v2}, we show the phase diagram for $k=1$, which is
calculated by using the function $V^{\prime \prime \prime }$. One can see
that the stable region in the low $\hat{\tau}$, near $R_{0}=1$, is changed
into a unconfirmed region. This behavior is commonly seen in the results
obtained using other three Lyapunov functions. As $k$ increases, this
propensity becomes prominent and the stable region shrinks increasingly. The
trajectories of the scaling solution are plotted in the same figure, again.
One can see that the trajectory of the scaling solution with $a=1$ passes
the unconfirmed region, although the trajectory with $a=0.1$ still stays
inside the stable region. It should be noted that, for $k=0$, at least a
part of the unconfirmed region near $R_{0}=1$ should be unstable. This
suggests that the scaling solution can become more unstable for the
inhomogeneous perturbation as the bulk viscosity $a$ increases.

\begin{figure}[tbp]
\begin{minipage}{.45\linewidth}
\includegraphics[scale=0.3]{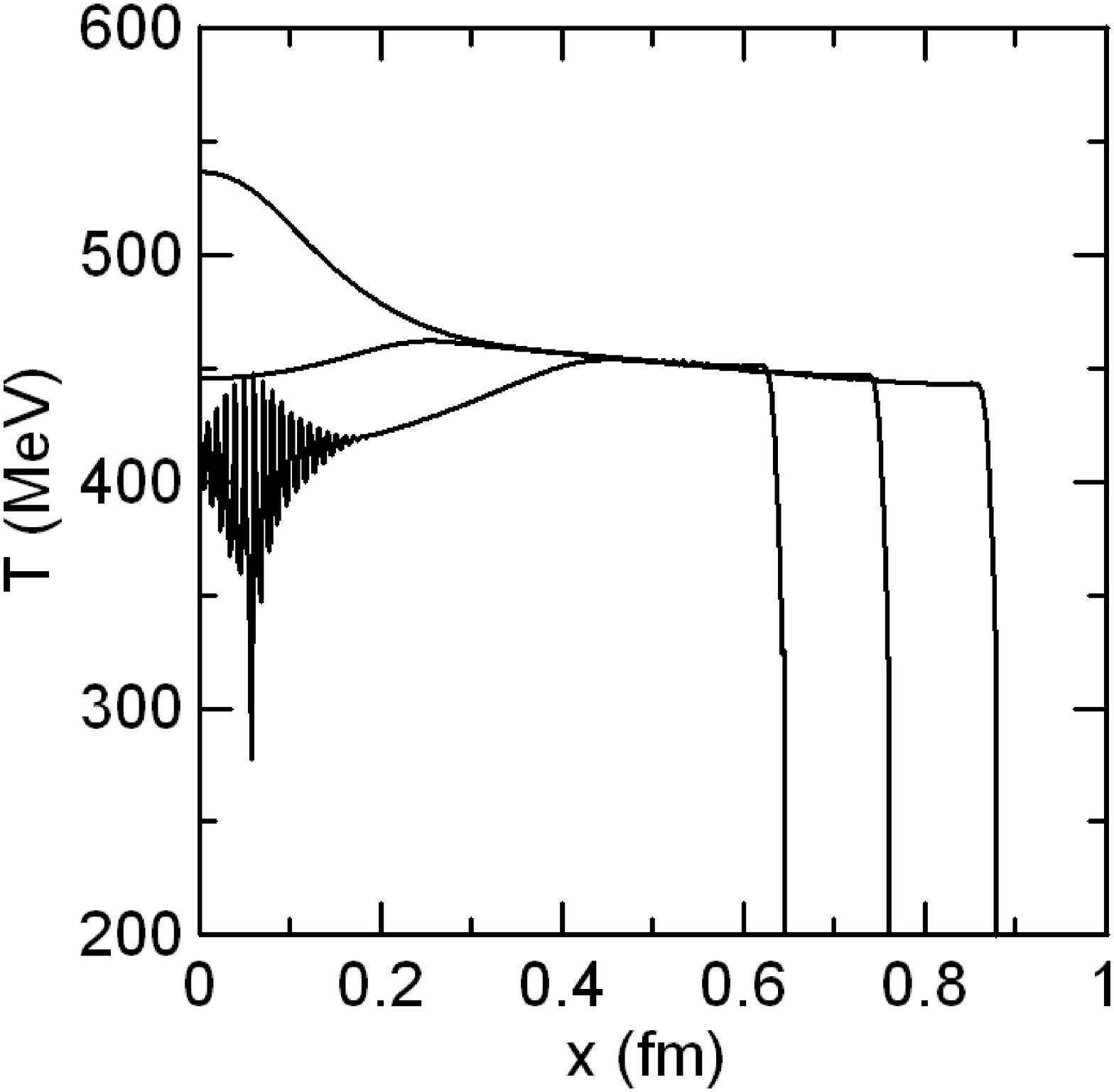}
\caption{The numerical simulation of the 1+1 dimensional CD hydrodynamics
with $a=1$ for $t=0.52$ $0.72$ and $0.92$ fm. The non-periodic oscillations appear in the center of the fluid.}
\label{tur}
\end{minipage}
\begin{minipage}{.45\linewidth}
\includegraphics[scale=0.3]{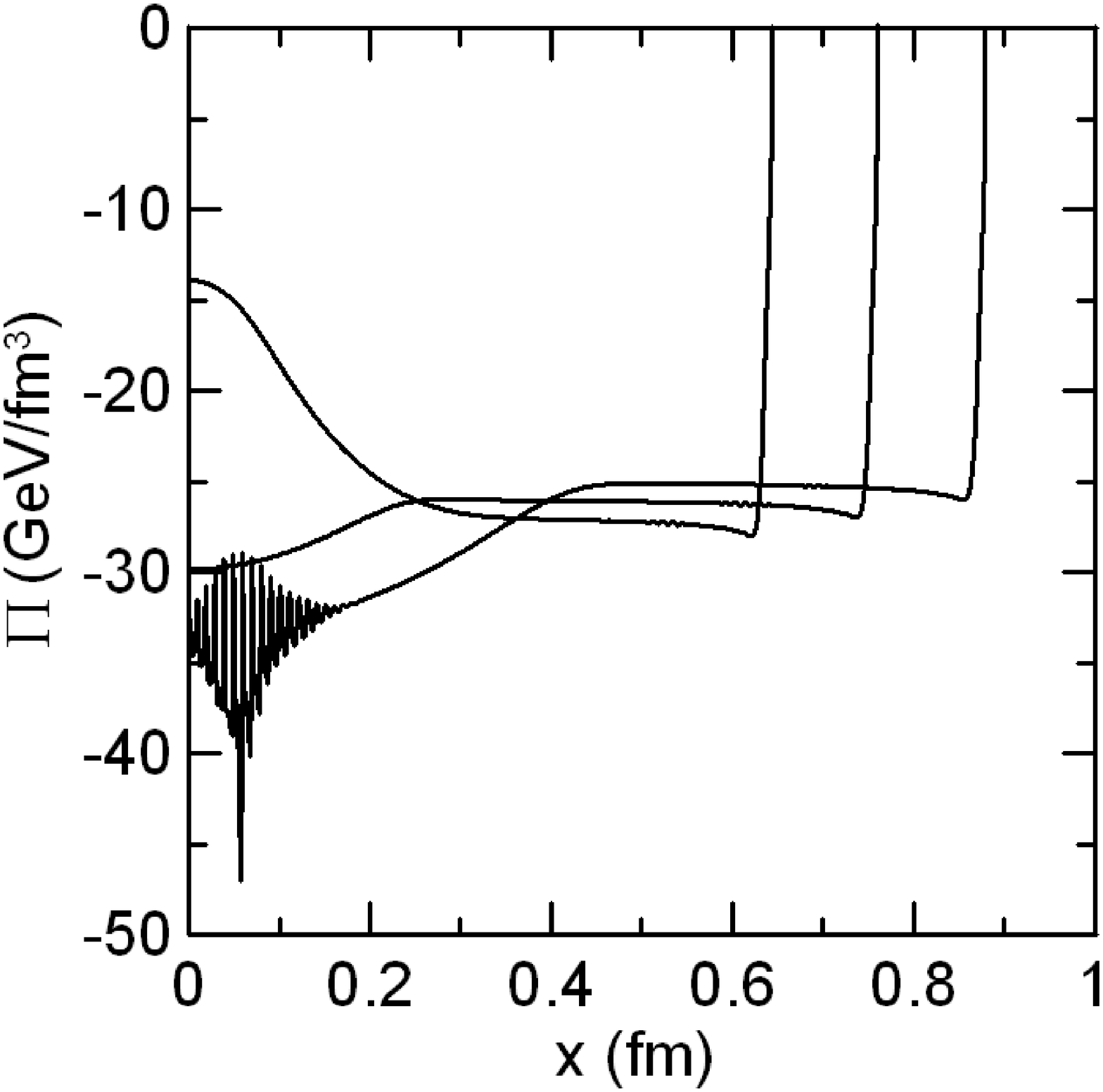}
\caption{The numerical simulation of the 1+1 dimensional CD hydrodynamics
with $a=1$ for $t=0.52$ $0.72$ and $0.92$ fm. The non-periodic oscillations appear in the center of the fluid.}
\label{tur_bulk}
\end{minipage}
\end{figure}

The situation discussed here seems to be realized in the numerical
simulations of the 1+1 dimensional CD hydrodynamics. When we increases the
bulk viscosity $a$, we found that the numerical calculation becomes unstable
and a kind of non-periodic oscillation appears in the center of the fluid.
In Figs. \ref{tur} and \ref{tur_bulk}, we show the temperature and the bulk
viscosity of the viscous fluid with $a=1$ for $t=0.52$ $0.72$ and $0.92$ fm,
respectively. We used the Landau initial condition with the initial
temperature $T=590$ MeV and the initial size $0.7$ fm. To remove numerical
oscillations which will disappear in the continuous limit, we used the
additional viscosity which is introduced in \cite{dkkm2}. It seems that the
oscillation appears when $R_{0}^{-1}$ exceeds a critical value by decreasing
the temperature and the bulk viscosity.  The amplitude of the
oscillation grows up with time and finally the numerical calculation
collapses.

So far, we discussed fixing the parameter $b=6$ which is consistent with
causality. Even when we use the acausal parameter set, $b=1$, quantitative
behavior of the phase diagram is not changed, but the stable region becomes
much smaller.

\section{Concluding remarks}

\label{chap:5}

Although the physical importance is recognized for the application
in QGP physics, the relativistic viscous hydrodynamics is not well
established yet. Some authors use the first order theory to estimate the
effect of viscosity in collective observables such as $v_{2}$
hoping that the deviation from the ideal hydrodynamics is small so that the
theory is of the first order or the second order might be irrelevant.
However, the difference between them might be fatal when any instabilities
or singularities emerge, such as shock wave propagations. Therefore, it is
fundamental to understand the stability of these theories. In this paper,
we discussed the causality and stability of the two cases of relativistic
dissipative hydrodynamics, the LL theory and CD hydrodynamics.

\begin{center}
\begin{tabular}{c|c|c|c}
& LL theory (acausal) & CD hydrodynamics (acausal) & CD
hydrodynamics (causal) \\ \hline
hydrostatic state & stable & stable & stable \\ 
moving frame & unstable & unstable & stable \\ 
scaling solution & stable/unstable & stable/unstable & stable/unstable%
\end{tabular}
\label{table}
\end{center}

The LL theory is known to be acausal whereas the CD hydrodynamics can be causal
depending on the values of parameters of the theory. 
The stability of the theories are summarized in the above Table. 
Around the hydrostatic
state, the LL theory and the CD hydrodynamics are stable. However, when we
move to a Lorentz boosted frame, the acausal theories like the LL theory and
the CD hydrodynamics with acausal parameter set become unstable. 
The second line shows that causality and stability are intimately
correlated in relativistic dissipative hydrodynamics. The stability of a
theory should not depend on the choice of frames. In this sense, the LL
theory and the CD hydrodynamics with acausal parameter sets are inconsistent.

The stability of the scaling solution was analyzed by using the Lyapunov
direct method. In the LL theory, it is known that the scaling solution is
stable against homogeneous and inhomogeneous perturbations when we use
initial conditions which satisfies $R_{0}\geq 1$ \cite{kouno}. In the CD
hydrodynamics, we found that the scaling solution cannot be stable even for $%
R_{0}\geq 1$. For the homogeneous perturbation ($k=0$), we confirmed that
most parts of the trajectories of the scaling solutions pass the stable
region in the phase diagram, which was plotted in terms of $\hat{\tau}$ and $%
R_{0}^{-1}$. Thus, the scaling solution will be stable for the homogeneous
perturbation. However, as the $k$ increases, the confirmed stable region in
the phase diagram shrinks and the trajectories of the scaling solutions
start to penetrate the unconfirmed region. When the unconfirmed regions are
real unstable regions, it means that the scaling solution is unstable for
inhomogeneous perturbations. This instability is distinguished for larger
bulk viscosity because the trajectory with larger viscosity is easier to
penetrate the unstable region.

Above conclusion may be supported by the numerical calculations. As a matter
of fact, we found that the numerical calculations of the 1+1 dimensional CD
hydrodynamics becomes unstable as the bulk viscosity coefficient increases
and a kind of non-periodic oscillations appears in the central rapidity
region. To see the quantitative signature of the oscillations, we have to
investigate various cases with different parameters and initial conditions.%
This oscillation may be interpreted as turbulence because the
instability of the scaling solution indicates chaos which acts as the
trigger of turbulence. However, in this work, we could not confirm the
unstable region on the phase diagram. To see the appearance of turbulence,
we need more systematic study of the instability around the scaling solution
beyond the Lyapunov direct method. 
This is a challenge for the future.

When we discuss the shear viscosity, 
we have to find the parameters which is consisitent with causality as was discussed 
in this paper.
This is also a future task.

\vspace{1cm} T. Koide acknowledges helpful discussions and comments with H.
Kouno, I. Mishustin, F. Takagi and G. Torrieri. This work is supported by
CNPq and FAPERJ.

\appendix 

\section{non-linear effect for causality}

\label{app:nonlinear}

The propagation speed of the fluid has been discussed based on the linear
analysis, and hence the effect of nonlinearity is ignored. In this section,
following the discussion of \cite{his5,his6}, we derive the effect of the
nonlinearity in the propagation speed.

For the simple 1+1 dimensional system, the hydrodynamic equations are
summarized as follows; 
\begin{eqnarray}
(A^{\mu}_{\nu})^t \partial_t Y^{\nu} + (A^{\mu}_{\nu})^x \partial_x Y^{\nu}
+ B^{\mu} =0,
\end{eqnarray}
where $Y^{\mu} =(\varepsilon,\theta,\Pi)$ and 
\begin{eqnarray}
(A)^t = \left( 
\begin{array}{ccc}
\cosh^2 \theta + (\cosh^2 \theta -1) \alpha & 2w \cosh \theta \sinh \theta & 
\sinh^2 \theta \\ 
\cosh \theta \sinh \theta (1+\alpha) & w(\cosh^2 \theta + \sinh^2 \theta) & 
\cosh \theta \sinh \theta \\ 
0 & \zeta \sinh \theta & \tau_R \cosh \theta%
\end{array}
\right),
\end{eqnarray}
\begin{eqnarray}
(A)^x = \left( 
\begin{array}{ccc}
\cosh \theta \sinh \theta (1+\alpha) & w(\cosh^2 \theta + \sinh^2 \theta) & 
\cosh \theta \sinh \theta \\ 
\sinh^2 \theta + (\sinh^2 \theta +1) \alpha & 2w \cosh \theta \sinh \theta & 
\cosh^2 \theta \\ 
0 & \zeta \cosh \theta & \tau_R \sinh \theta%
\end{array}
\right),
\end{eqnarray}
\begin{eqnarray}
B^{\mu} = (0,0,0,\Pi).
\end{eqnarray}
Here, $w$ is the effective enthalpy density, $w=\varepsilon + P + \Pi$.

The characteristic speed $v$ is given by 
\begin{eqnarray}
det(v(A)^t - (A)^x)=0.
\end{eqnarray}
The speed is easily estimated in the local rest frame, $\theta=0$. Then we
have the following three solutions, 
\begin{eqnarray}
v= 0, \pm\sqrt{\frac{\alpha w \tau_R + \zeta}{w\tau_R}}.
\end{eqnarray}
One can easily see that if $\Pi$ is small and we can replace $w$ with $%
\varepsilon + P$, this result is same as Eq. (\ref{eqn:Lk}).

We consider the case of the effective enthalpy density is positive. 
Then, to satisfy causality, the transport coefficients should
satisfy the following condition, 
\begin{eqnarray}
\frac{\zeta}{\tau_R} \le (1-\alpha) (\varepsilon + P + \Pi).
\end{eqnarray}
This is, again, the generalization of the restriction for the transport
coefficients discussed below Eq. (\ref{eqn:def-tau}).

\section{instability in general equilibrium frame (Hiscock-Lindblam)}

\label{app:his}

In section \label{chap:move}, the stability from a Lorentz boosted frame was
discussed. A similar problem was discussed by Hiscock and Lindblam \cite%
{his3}. In this appendix, we apply their discussion to the CD hydrodynamics.

They consider the transformation of the coordinate by using the following
replacement of the variables, 
\begin{eqnarray}
\omega &=& \gamma(\tilde{\omega} + v \tilde{k}), \\
k &=& \gamma(\tilde{k} + v \tilde{\omega}),
\end{eqnarray}
where $\omega$ and $k$ are variables in the rest frame, and $\tilde{\omega}$
and $\tilde{k}$ are in the new frame, which moves with the velocity $v$.
Substituting into the result obtained in the rest frame (\ref{eqn:disp}), we
have 
\begin{eqnarray}
\gamma^3 (\tilde{\omega} + v \tilde{k})^3 - \frac{i}{\tau_R}\gamma^2 ( 
\tilde{\omega} + v \tilde{k} )^2 - \left( \frac{\zeta}{\tau_R}\frac{1}{%
\varepsilon + P}+ \alpha \right) \gamma^3 (\tilde{k} + v \tilde{\omega})^2 (%
\tilde{\omega} + v \tilde{k}) + i \frac{\alpha}{\tau_R}\gamma^2 (\tilde{k} +
v \tilde{\omega})^2 =0 .
\end{eqnarray}
We can easily solve the equation for $\tilde{k}=0$, 
\begin{eqnarray}
\tilde{\omega} = 0,0, \frac{i}{\tau_R}\frac{1-\alpha v^2}{\gamma (1-\frac{%
\zeta}{\tau_R} \frac{v^2}{\varepsilon + P} - \alpha v^2)}.
\end{eqnarray}
The imaginary part is positive and hence the theory is still stable.

On the other hand, in the LL theory, the solutions are given by 
\begin{eqnarray}
\tilde{\omega} = 0, 0, -i \frac{(\varepsilon + P)(1-\alpha v^2)}{\zeta
\gamma v^2}.
\end{eqnarray}
Thus, the LL theory is unstable again.

\section{Lyapunov direct method}

\label{app:sta}

Here, we summarize the stability analysis based on the Lyapunov function. As
an example, let us consider the damped harmonic oscillator, 
\begin{eqnarray}
\frac{\partial}{\partial t}x &=& v, \\
\frac{\partial}{\partial t}v &=& -\gamma v - \omega^2 x .
\end{eqnarray}
The solution of the equation converges to $x=v=0$.

To discuss the stability around the equilibrium solution $x_0=v_0=0$, we
introduce a function which characterizes the deviation from the
equilibrium. For example, we choose 
\begin{eqnarray}
V = (v-v_0)^2 + \alpha^2 (x-x_0)^2.  \label{eqn:func1}
\end{eqnarray}
This is positive definite and if this function monotonically decreases with
time, the system is stable and the function $V$ is called the Lyapunov function. 
The time evolution of the function $V$ is given by 
\begin{eqnarray}
\frac{d}{dt}V = (\alpha (x-x_0), v-v_0) M \left( 
\begin{array}{c}
\alpha (x-x_0) \\ 
v-v_0%
\end{array}
\right),
\end{eqnarray}
where 
\begin{eqnarray}
M = \left( 
\begin{array}{cc}
0 & \alpha - \frac{\omega^2}{\alpha} \\ 
\alpha - \frac{\omega^2}{\alpha} & -2\gamma%
\end{array}
\right).
\end{eqnarray}
The eigen values of the matrix $M$ are given by 
\begin{eqnarray}
\lambda_{\pm} = \frac{1}{2}(-\gamma \pm \sqrt{\gamma^2 +
(\alpha-\omega^2/\alpha)^2}).
\end{eqnarray}
One can see that when $\alpha= \omega$, $V$ is a monotonically decreasing function in time.
Thus $V$ is the Lyapunov function and the equilibrium state is stable.

However, if we use $\alpha \neq \omega$, the function generally 
have a positive and negative eigen values.
Thus we cannot determine the stability of the equilibrium solution.
In this sense, the Lyapunov direct method usually underestimates the stability 
of the system.

Similarly, when we find
that the minimum eigen value is positive 
and hence $V$ is a monotonically increasing function, the equilibrium solution is
unstable.

\section{Another case of the function $V^{\prime\prime\prime}$}

\label{app:V}

Instead of Eq. (\ref{eqn:X}), we introduce the following vector, 
\begin{eqnarray}
X = \left( 
\begin{array}{c}
\delta \phi \\ 
\delta \theta \\ 
R^{-1}_0 \delta \ln \Pi%
\end{array}
\right).
\end{eqnarray}
Then, the Lyapunov function is given by 
\begin{eqnarray}
V^{\prime\prime\prime}= X^{\dagger}X = |\delta \phi|^2 + |\delta \theta|^2 +
(R^{-1}_0)^2 |\delta \ln \Pi|^2 .
\end{eqnarray}
Then the evolution equation of the Lyapunov function is 
\begin{eqnarray}
\tau \partial_{\tau}V^{\prime\prime\prime}= X^{\dagger} (A^{\dagger} + A) X ,
\end{eqnarray}
where 
\begin{eqnarray}
A = \left( 
\begin{array}{ccc}
- (1-R^{-1}_0) \left( \frac{\partial \sqrt{\alpha}}{\partial \phi} \right)_0
- R^{-1}_0 \left( 1 + \alpha \right) & - (1-R^{-1}_0) \sqrt{\alpha} (-ik) & 
\sqrt{\alpha} \\ 
- \sqrt{\alpha} \frac{(-ik)}{(1-R^{-1}_0)} & - (1-\alpha) - \frac{(R^{-1}_0 
\hat{\tau} - \frac{1}{b} )}{(1-R^{-1}_0)} & \frac{(-ik)}{(1-R^{-1}_0)} \\ 
- \sqrt{\alpha} \left( R^{-1}_0 \hat{\tau} - \frac{1}{ b}\right) + \frac{1}{%
\sqrt{\alpha} b} & \frac{-ik}{ b} & - \hat{\tau} - (1 - R^{-1}_0)(1+\alpha)%
\end{array}
\right).
\end{eqnarray}

Similarly, as for $R_0 = 1$, the matrix $B$ is given by 
\begin{eqnarray}
B = \left( 
\begin{array}{ccc}
- \left( 1 + \alpha \right) & \sqrt{\alpha} &  \\ 
- \sqrt{\alpha} \left( \hat{\tau} - \frac{1}{b} \right) + \frac{1}{\sqrt{%
\alpha}b} + \frac{\sqrt{\alpha} k^2}{b \hat{\tau} -1} & - \hat{\tau} -\frac{%
k^2}{ b \hat{\tau}-1} & 
\end{array}
\right).
\end{eqnarray}

\end{document}